\documentclass[10pt,preprint]{aastex}
\newcommand{\RXTE}{\emph{Rossi \xray\ Timing Explorer}}
\newcommand{\rxte}{\emph{RXTE}}

\newcommand{\asm}{ASM}
\newcommand{\PCA}{Proportional Counter Array}
\newcommand{\pca}{PCA}

\newcommand{\PCUS}{Proportional Counter Units}
\newcommand{\pcus}{PCUs}
\newcommand{\HEXTE}{High Energy \xray\ Timing Experiment}
\newcommand{\hexte}{HEXTE}

\newcommand{\sax}{\emph{Beppo}SAX}

\newcommand{\rosat}{\emph{ROSAT}}
\newcommand{\uhuru}{\emph{UHURU}}
\newcommand{\arielv}{\emph{Ariel V}}

\newcommand{\tenma}{\emph{Tenma}}
\newcommand{\ginga}{\emph{Ginga}}

\newcommand{\HEXE}{\emph{High Energy \xray\ Experiment}}
\newcommand{\hexe}{HEXE}

\newcommand{\copernicus}{\emph{Copernicus}}

\newcommand{\heao}{\emph{HEAO}}

\newcommand{\bbxrt}{\emph{BBXRT}}

\newcommand{\cgro}{\emph{CGRO}}

\newcommand{\batse}{BATSE}

\newcommand{\osse}{OSSE}

\newcommand{\CRSFs}{Cyclotron Resonance Scattering Features}
\newcommand{\CRSFS}{Cyclotron Resonance Scattering Features}
\newcommand{\crsf}{CRSF}
\newcommand{\crsfs}{CRSFs}

\newcommand{\xray}{X-ray}
\newcommand{\xrays}{X-rays}
\newcommand{\gammaray}{\ensuremath{\gamma{\rm -ray}}}

\newcommand{\faint}{Faint}
\newcommand{\skyvle}{SkyVLE}

\newcommand{\npex}{NPEX}

\newcommand{\fdco}{FDCO}

\newcommand{\plcut}{PLCUT}
\newcommand{\mplcut}{MPLCUT}

\newcommand{\gabs}{GABS}

\newcommand{\hmxb}{HMXB}

\newcommand{\lmxb}{LMXB}

\newcommand{\bet}{Be-Transient}

\newcommand{\bep}{Be-Persistent}

\newcommand{\tuc}{Transient}

\newcommand{\heasarc}{HEASARC}
\newcommand{\heasoft}{\texttt{HEASoft}}
\newcommand{\ftool}{\texttt{FTOOL}}
\newcommand{\ftools}{\texttt{FTOOLs}}
\newcommand{\xanadu}{\texttt{XANADU}}

\newcommand{\xspec}{\texttt{XSPEC}}
\newcommand{\ftest}{F-Test}

\newcommand{\pcmsq}{\ensuremath{\rm{cm^{-2}}}}
\newcommand{\flux}{\ensuremath{\rm{ergs\,cm^{-2}\,s^{-1}}}}
\newcommand{\lum}{\ensuremath{\rm{ergs\,s^{-1}}}}

\newcommand{\cpspcu}{\ensuremath{\rm{cts\,s^{-1}\,PCU^{-1}}}}
\newcommand{\wsim}{\ensuremath{\sim}}

\newcommand{\wpm}{\ensuremath{\pm}}
\newcommand{\wchi}{\ensuremath{\chi^{2}}}
\newcommand{\wchired}{\ensuremath{\chi^{2}_{\rm{red}}}}

\newcommand{\wgamma}{\ensuremath{\Gamma}}
\newcommand{\nh}{\ensuremath{N_{\rm{H}}}}
\newcommand{\ecut}{\ensuremath{E_{\rm{cut}}}}
\newcommand{\efold}{\ensuremath{E_{\rm{fold}}}}
\newcommand{\ecy}[1]{\ensuremath{E_{\rm{c#1}}}}
\newcommand{\wcy}[1]{\ensuremath{\sigma_{\rm{c#1}}}}
\newcommand{\dcy}[1]{\ensuremath{\tau_{\rm{c#1}}}}

\newcommand{\kte}{\ensuremath{kT_{\rm e}}}

\newcommand{\rns}{\ensuremath{R_{\rm ns}}}
\newcommand{\mns}{\ensuremath{M_{\rm ns}}}
\newcommand{\msol}{\ensuremath{\rm M_{\odot}}}
\newcommand{\porb}{\ensuremath{P_{\rm orb}}}
\newcommand{\pspin}{\ensuremath{P_{\rm spin}}}

\newcommand{\degree}{\ensuremath{^\circ}}

\newcommand{\mbf}[1]{\ensuremath{\mathbf{#1}}}

\newcommand{\apjplcuteq}{\ensuremath{
   \mathrm{PLCUT}(E) = A\ E^{-\Gamma}\times\
   \cases{1&$(E\leq\ecut)$ \cr {\rm e}^{-(E-\ecut)/\efold}&$(E>\ecut)$}
}}

\newcommand{\fdcoeq}{\ensuremath{
   \mathrm{FDCO}(E) = A\ E^{-\Gamma}\
   \frac{1}{1 + \mathrm{e}^{(E-\ecut)/\efold}}
}}

\newcommand{\npexeq}{\ensuremath{
   \mathrm{NPEX}(E) = A(E^{-\Gamma_{1}} + BE^{+\Gamma_{2}})
   \mathrm{e}^{-E/\efold}
}}

\newcommand{\crsfeq}[1]{\ensuremath{
   \ecy{#1}= 11.6\frac{B}{10^{12}\,\rm{G}}(1 + z)^{-1}
}}

\newcommand{\revcrsfeq}[1]{\ensuremath{
   \ecy{#1} = m_{\rm e}c^{2} \frac
   {(1 + 2nB'\sin^{2}(\theta))^{1/2} - 1}
   {\sin^{2}(\theta)}
}}

\newcommand{\gabseq}[1]{\ensuremath{
   {\rm GABS}(E) = \dcy{#1}\
   {\rm e}^{-(E-\ecy{#1})^{2}/(2\wcy{#1}^{2})}
}}

\newcommand{\taueq}{\ensuremath{
   I_{0}(E) \rightarrow I_{0}(E) e^{\rm{-GABS}(E)}
}}

\newcommand{\eqfwhmpropto}{\ensuremath{
   \Gamma_{\rm c} \propto \ecy{} \sqrt{\kte} \left|\cos(\theta)\right|
}}

\begin{document}

%% /*******************************************************************
%% ** The Header                                                     **
%% *******************************************************************/

\title{Magnetic Fields of Accreting X--Ray Pulsars with the
\emph{Rossi X--Ray Timing Explorer}}

\author{W. Coburn\altaffilmark{1}$^{,}$\altaffilmark{2},
 W. A. Heindl\altaffilmark{2},
 R. E. Rothschild\altaffilmark{2},
 D. E. Gruber\altaffilmark{2},
 I. Kreykenbohm\altaffilmark{3},
 J. Wilms\altaffilmark{3},
 P. Kretschmar\altaffilmark{4}$^{,}$\altaffilmark{5},
 R. Staubert\altaffilmark{3}
}

\altaffiltext{1}{Space Sciences Laboratory,
University of California at Berkeley, Berkeley, CA, 94702-7450, USA}

\altaffiltext{2}{Center for Astrophysics and Space Sciences,
University of California at San Diego, La Jolla, CA,
92093-0424, USA}

\altaffiltext{3}{Institut f\"{u}r Astronomie und Astrophysik,
Astronomie, University of T\"{u}bingen, Sand 1, D-72076, T\"{u}bingen,
Germany}

\altaffiltext{4}{INTEGRAL Science Data Center, 6 ch. d'\'Ecogia,
CH-1290 Versoix, Switzerland}

\altaffiltext{5}{Max-Planck-Institut f\"ur Extraterrestrische
Physik, Giessenbachstrasse 1, D-85740 Garching, Germany}

\setcounter{footnote}{0}

%% /*******************************************************************
%% ** The Abstract                                                   **
%% *******************************************************************/

\begin{abstract}

Using a consistent set of models, we parameterized the \xray\ spectra
of all accreting pulsars in the \RXTE\ database which exhibit \CRSFS\
(\crsfs, or cyclotron lines). These sources in our sample are Her X-1,
4U 0115+63, Cen X-3, 4U 1626-67, XTE J1946-274, Vela X-1, 4U 1907+09,
4U 1538-52, GX 301-2, and 4U 0352+309 (X Per). We searched for
correlations among the spectral parameters, concentrating on how the
cyclotron line energy relates to the continuum and therefore how the
neutron star $B$-field influences the X-Ray emission. As expected, we
found a correlation between the \crsf\ energy and the spectral cutoff
energy. However, with our consistent set of fits we found that the
relationship is more complex than what has been reported
previously. Also, we found that not only does the width of the
cyclotron line correlate with the energy (as suggested by theory), but
that the width scaled by the energy correlates with the depth of the
feature. We discuss the implications of these results, including the
possibility that accretion directly affects the relative alignment of
the neutron star spin and dipole axes. Lastly, we comment on the
current state of fitting phenomenological models to spectra in the
\rxte/\sax\ era and the need for better theoretical models of the
\xray\ continua of accreting pulsars.

\end{abstract}


%% /*******************************************************************
%% ** Introduction                                                   **
%% *******************************************************************/

\section{Introduction}\label{sec:intro}

Accretion powered \xray\ pulsars \citep{whi83,nag89,bil97} provide a
unique laboratory for the study of matter in extremes of temperature
and magnetic as well as gravitational fields. After more than two
decades of research, however, there is still no compelling model for
the generation of the hard \xray\ spectrum in these objects. This
reflects the difficulties and complexities of radiative transport and
magnetohydrodynamics in the environment found at the neutron star
magnetic polar caps.

The observed hard \xray\ emission from these objects originates
primarily from one or two ``hotspots'' found at the neutron star
magnetic poles. Due to their large magnetic fields
($B\gtrsim10^{12}$\,G), material accreted from a nearby companion
couples to the neutron star $B$-field at several hundred neutron star
radii. The material is then channeled onto the neutron star surface,
forming accretion structures at the two poles. It is the combination
of the beaming properties of these structures with the rotation of the
star that gives rise to the pulsed emission seen by a distant
observer. By analyzing the rotation-averaged spectral properties of
these hotspots, we hope to improve our understanding of the physical
conditions and properties of the emission regions of these accretion
structures. This is also a step towards interpreting and understanding
pulsar spectra as a function of neutron star rotation phase (pulse
phase resolved spectroscopy).

For the analysis presented here, we focus on the effects of the
magnetic field on the resulting hard \xray\ spectrum. By using sources
with known magnetic field strengths (from the measurement of cyclotron
features) we remove one uncertainty from the class analysis. Our
results, however, should also extrapolate to accreting pulsars with
unknown $B$-field strengths. Another motivation was to provide an
observational base for theoretical investigations into the production
of the pulsar hard \xray\ continuum, as well as to guide future
calculations and simulations.  The list of our sources, along with
some of their properties, is given in Table~\ref{table:systems}.

To perform this analysis, we used sources where there was a direct
measurement of the pulsar $B$-field using \CRSFs\ (\crsfs), also
commonly referred to as ``cyclotron lines.'' These line-like spectral
features arise due to the resonant scattering of photons by electrons
whose energies are quantized into Landau levels by the magnetic field
\citep{meszaros92}. The fundamental energy where these features appear
is given by
\begin{equation}\label{eq:crsfeq}
\crsfeq{}\rm{\,keV}
\end{equation}
where $B$ is the magnetic field (in Gauss) in the scattering region,
and $z$ is the gravitational redshift. The quantized energy levels of
the electrons are to first order harmonically spaced, with features at
2\ecy{}, 3\ecy{}, etc. both predicted and, in some sources,
observed \citep[e.g.][]{hei99comp0115,san99,cus98}. At sufficiently
high magnetic fields, relativistic effects can introduce a slight
anharmonicity in the rest frame resonant photon energies. In these
cases the cyclotron energies are given by \citep{har91}
\begin{equation}\label{eq:revcrsfeq}
\revcrsfeq{}
\end{equation}
where $B'=B (\hbar e)/(m_{e}^{2}c^{3})$ is the magnetic field scaled
to the QED field scale, $n$ is the harmonic number, and $\theta$ is
the angle of propagation of the photon relative to the magnetic
field. Since these energies depend on the angle $\theta$, the emergent
spectral features are influenced heavily by the spatial distribution
of electrons in the scattering region.

The dependence of $\theta$ in Eq.~\ref{eq:revcrsfeq} indicates that,
even at nonrelativistic energies, the \crsf\ energy is not the only
source of information about the scattering region. From the Monte
Carlo simulations of \citet{ara99} and \citet{ise98,ise98b}, it is
found that the shape of the fundamental can be quite complex, and in
general depends heavily on the details of the emission and scattering
geometries, as well as the physical parameters such as the electron
temperature and density in the scattering region. These features are
also sometimes observed to vary as a function of rotation phase of the
star \citep[e.g.][]{bur00,biff99,cla90,soo90a,vog82}, allowing for the
detailed study of a single accretion structure using pulse phase
resolved spectroscopy.

In this paper we summarize spectral fits to ten accreting pulsars, and
present observational evidence for the effect of the $B$-fields on the
underlying hard \xray\ continua of these pulsars. This was part of a
larger analysis of \rxte\ archival data that encompassed 25 accreting
pulsars in total. These ten (see Table~\ref{table:systems}) were
selected due to the fact that their spectra exhibited \crsfs. They
represent a complete sample of pulsars with \crsfs\ in the \rxte\
database. The remaining sources, the ones without detectable cyclotron
lines, will be discussed in a future publication.

In \S\ref{sec:rxte} we discuss the \RXTE\ (\rxte) satellite. In
\S\ref{sec:method} we present a summary of the methodology and
spectral models we used. In \S~\ref{sec:obs} we discuss the 10 pulsars
in our sample, along with fits to their \rxte\ spectra. In
\S\ref{sec:fits} the results of the fitting are presented, along with
a discussion of how correlations were found and the checks that were
done using Monte Carlo simulations. In \S\ref{subsec:results} we
discuss our findings and their physical implications. Finally, in
\S\ref{sec:summary} we conclude with a brief summary of our primary
results and discoveries.

%% /*******************************************************************
%% ** The RXTE                                                       **
%% *******************************************************************/

\section{The \rxte}\label{sec:rxte}

All of our observations here were made using the two pointed
instruments on the \rxte. The \PCA\ \citep[\pca,][]{jah96} is a set of
five nearly identical Xenon proportional counters (\PCUS, or \pcus)
sensitive in the energy range 2--60\,keV. Each detector has a
geometric area of \wsim1300\,cm\,$^2$ and an energy resolution of 18\%
at 6\,keV. To extend the lifetime of the instrument, the \pca\ often
operates with one or more \pcus\ turned off, resulting in a reduced
effective area during those observations. The \HEXTE\
\citep[\hexte,][]{rot98} consists of two clusters of 4 NaI(Tl)/CsI(Na)
phoswich scintillation detectors (15--250\,keV) that rock on and off
source, alternately measuring source plus background and background
flux. The physical area of each cluster is \wsim800\,cm$^2$, and the
energy resolution is 16\% at 60\,keV. A pulse height analyzer in one
of the detectors in Cluster B failed early in the mission and no
longer provides any spectral information, effectively reducing the
area of that cluster by 25\%. Observations done before the failure are
noted in the text. The \pca\ and \hexte\ fields of view are co-aligned
on-source and are collimated to the same 1\degree\ full width half
maximum (FWHM) region.

The high voltage of the \pca\ has been changed several times during
the mission, resulting in four ``Gain Epochs'' \citep{jah96}.  The
spectra presented here are primarily from \pca\ epoch 3, although
there are two observations that were done during epoch 1 (see
Table~\ref{table:pcaobs}). Due to the relatively short duration of
epoch 1, there has been less emphasis on understanding the calibration
and producing background models during these periods. Therefore the
observations obtained during this time have larger systematic
uncertainties associated with them when compared to epoch 3
observations (see below). The gain of the \hexte\ detectors has been
held fixed by the automatic gain control system \citep{pel91} since
launch, and therefore there is no \hexte\ analogy to the \pca\ gain
epochs.

\pca\ backgrounds were estimated using one of 2 background
models released by the \pca\ instrument team, the choice of which
depends on the source counting rate. The FAINT model is for source
count rates below 40\,\cpspcu, while for brighter sources the
\skyvle\ model was used. Comparisons of the simulated and measured
background spectra and light curves show that these models work
well, especially below \wsim20\,keV where our \pca\ analysis is
concentrated \citep{jah96}. As mentioned above, the \hexte\ clusters
alternate pointing between target and nearby blank fields in order to
measure, rather than model, the background.

The uncertainties in the \pca\ spectra, which have excellent counting
statistics, are dominated by the systematic errors in the response
matrix and background modeling \citep{wil99}. To estimate the size of
the systematic errors in the \pca\ version 2.43 response matrix, we
analyzed observations of the Crab nebula and pulsar as recommended by
the \pca\ team (Keith Jahoda, private communication). To fit the Crab
we used a combination of two power-laws; the first power-law was to
account for the nebula flux while the second models emission from the
pulsar. The 2--10\,keV normalization of the second power-law was fixed
to be 10\% of the first \citep{jah00,kni82}. We then examined the best
fit and increased the size of the systematic errors as a function of
energy until we achieved a reduced \wchi\ of unity. The second
power-law is important in this type of analysis. Ignoring the
contribution of the pulsar would have led us to estimate larger
systematic errors, and the power-law index obtained for the nebula
would disagree with what is observed by \hexte.

\begin{figure}[H]
\includegraphics{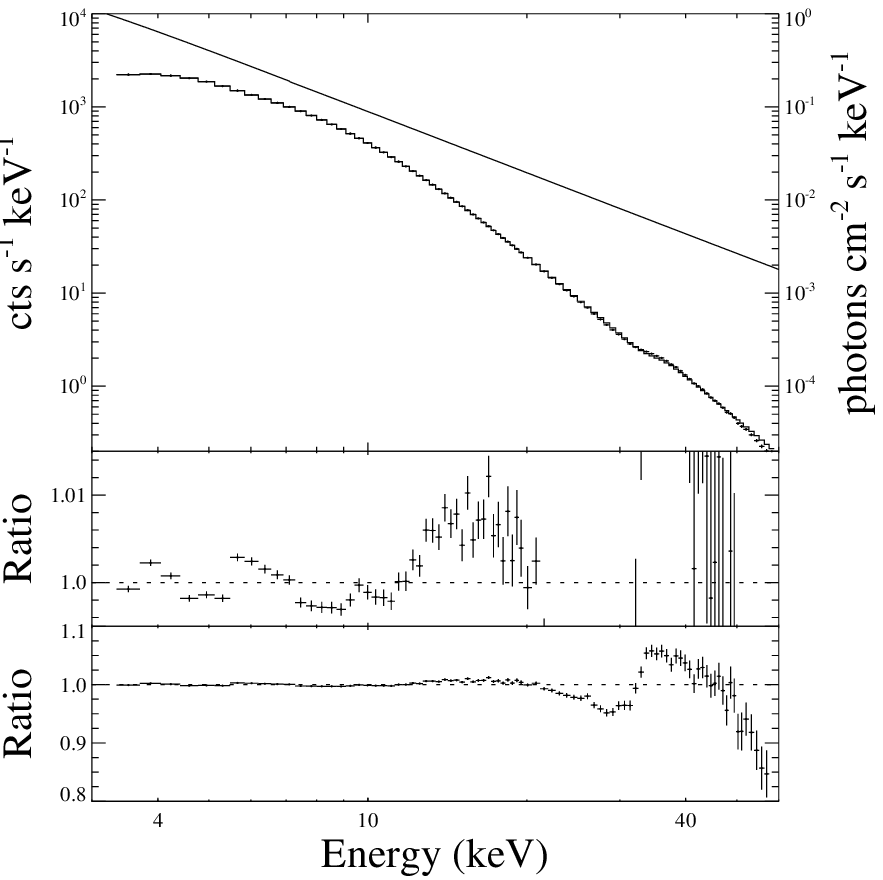}
\caption{\label{fig:pcacrab}
Top: A dual power-law fit to the \pca\ spectrum of the Crab
nebula/pulsar (see text). Shown are the counts spectrum (crosses),
inferred photon spectrum (smooth curve, see appropriate source section
for details) and model folded through the \pca\ response matrix
(histogram). The observations was obtained on 1996 August 23 and
lasted 10\,ks with all 5 \pcus. The fit is using two power-laws, the
first with an index of $2.245_{-0.003}^{+0.004}$, and the second with
an index of $1.87_{-0.02}^{+0.01}$ and a normalization constrained to
be 10\% that of the first power-law. Middle: Ratio of the data to the
best fit dual power-law model. Below \wsim12\,keV the size of the
residuals is approximately \wpm0.4\%. Bottom: Same plot as in the
middle pane, only with an enlarged vertical scale. Due to the large
line-like residual in the response matrix, we have limited our
analysis of \pca\ data to energies below 20\,keV.}
\end{figure}

When analyzing the results of fits to \pca\ data with no systematic
errors applied, there is a large line-like residual starting at
\wsim20\,keV due the K-edge of Xenon (see Fig.~\ref{fig:pcacrab}).
This is typically seen in all \pca\ spectra, and there are two methods
to account for it (Keith Jahoda, private communication). The first is
to use much larger systematic errors in this region, of order
5\%. This is necessary because of the size of the residual.  However,
since we are interested in searching for and analyzing spectral lines
in this region, we opted for the second method which ignores this part
of the \pca\ spectra completely to avoid misinterpretation. Below
20\,keV, the systematic errors used depend upon the gain epoch of the
\pca\ during the observation. For gain epoch 3, where the calibration
of the \pca\ is well understood, the typical systematic error at a
given energy below 20\,keV is 0.4\%. For gain epoch 1, which was
shorter and earlier in the mission, 0.8\% systematic errors were
required below 20\,keV. The application of systematic errors to the
data, although necessary, does mean that the use of the \wchi\
statistic in fitting is no longer strictly valid. Still, quoting
\wchi\ when doing spectral fitting is a good indication of how well a
model fits the data.

Unlike with the \pca, the dominant source of errors in spectra
obtained with \hexte\ are associated with photon counting
statistics. This is due partly because the collecting area of
\hexte\ is much less than that of the \pca, and partly because of
the lower intrinsic source fluxes in the \hexte\ energy range. To this
statistical limit the, \hexte\ calibration is reasonably well
understood and the released response matrices work well and agree with
the \pca\ in the range of energies where the two instruments overlap.
For our analysis we allowed for the overall normalization between the
two instruments to vary freely, although in all cases it was within
errors of what was expected.

We accumulated and analyzed the data using the standard NASA released
\heasoft\ package of tools released by the High Energy Astrophyics
Science Archive Research (\heasarc) Center\footnote{See
\underline{http://heasarc.gsfc.nasa.gov/} on the world wide web}. The
\heasoft\ consists of two packages, the \ftools\ data accumulation
software, and the the \xanadu\ spectral, timing, and image analysis
software. The entire process of accumulating data using the \ftools\
is outlined in ``The \rxte\ Cook Book: Recipies of Data Analysis and
Reduction,'' provided by the \heasarc\ on the \rxte\ Guest Observers
website.

When determining source ``goodtime'' intervals, times from which
to allow the extraction of source (or background) counts, we used
the following data selection criteria. To avoid possible
contamination from \xrays\ from the Earth's limb, data was
accumulated only when the satellite was pointing more than
10\degree\ above the horizon. To avoid possible contamination
from activation in the detectors due to the high particle rates
associated with SAA passages, we rejected data from a 30 minute
interval beginning with the satellite entering the SAA. We also
required that the satellite be pointing to within 0\degree.01 of
the source position. Lastly, for faint sources and with the \pca\
instrument only, the electron ratio in each of the accumulated
\pcus\ was required to be less than 0.10.

All other data accumulation steps, such as the modeling of the
\pca\ background, modeling \hexte\ deadtime corrections, and
generating response matrices, were done using the released \heasoft\
\ftool\ appropriate for the task. For spectral analysis we used the
\xspec\ fitting package, released as part of \xanadu\ in the
\heasoft\ tools. I initially used version 10.00, and later when
it was released version 11.0.1 \citep{arn96}.

%% /*******************************************************************
%% ** Methodology                                                    **
%% *******************************************************************/

\section{Methodology}\label{sec:method}

When analyzing lines in pulsar spectra, a good model of the underlying
continua is very important. Unfortunately, as mentioned in
\S\ref{sec:intro}, there exists no convincing theoretical model for
the shape of the continuum \xray\ spectrum in accreting \xray\ pulsars
\citep[][and references therein]{har94}. Therefore observers are
forced to use phenomenological models, with the choice of the specific
form left to the observer. In the \rxte\ band (2--250\,keV) these
models take the general form of a power law that changes into a
power-law times an exponential above a characteristic cutoff energy
\citep{whi83}, which we will refer to as either the standard \xray\
continuum shape or simply the standard pulsar spectrum.

There are three primary analytic realizations of the standard pulsar
continuum shape that have been used by various authors to fit \xray\
pulsar spectra. The first, and the one used in this paper, is a
power-law with a high-energy cutoff \citep[\plcut,][]{whi83}. The
analytic form of the \plcut\ model is:
\begin{equation}\label{eq:plcut}
\apjplcuteq
\end{equation}
where \wgamma\ is the photon index, and \ecut\ and \efold\ are the
cutoff and folding energies respectively. This form was chosen because
it was particularly suited for our parameterization analysis (see
below). The second is a power-law with a Fermi-Dirac form of the
cutoff
\citep[\fdco,][]{tan86}. Analytically this is given by the equation
\begin{equation}\label{eq:fdco}
\fdcoeq
\end{equation}
where $\Gamma$ is the photon index and \ecut, \efold\ the cutoff and
folding energies, respectively. The third consists of dual power-laws
with indices of opposite sign along with an exponential cutoff
\citep[\npex,][]{mih95}. This is realized as
\begin{equation}\label{eq:npex}
\npexeq
\end{equation}
where $\Gamma_{1}$ and $\Gamma_{2}$ are photon indices with positive
values, and \efold\ the folding energy. As can be readily seen from
Eqs.~\ref{eq:plcut}, \ref{eq:fdco}, and \ref{eq:npex}, even though the
models produce similar shapes, the details of the various functional
forms make a meaningful comparison of fit parameters across models
impossible. This underscores the need to use a single model
consistently for all of the spectra in order to make generalizations
about the sources as a class.

For the fits discussed in \S\ref{sec:obs} and presented in
Tables~\ref{table:plcutfits} and \ref{table:params}, we used a
Modified-\plcut\ model (\mplcut). The standard \plcut\ form
(Eq.~\ref{eq:plcut}) has some definite advantages over the other
models that are discussed below. Unfortunately, however, the slope of
the \plcut\ model has a discontinuity at the cutoff energy
\ecut, which can create an artificial line-like feature in the
spectral fit that can adversely affect fits to a \crsf\
\citep{kre97,bur00,bar01,coburn01}. To account for this, we applied a
``smoothing'' function in the form of a relatively narrow
($\sigma\lesssim0.1\ecy{}$), shallow ($\tau\lesssim0.1$) Gaussian
shaped absorption feature (\gabs, see below) at the cutoff
energy. While this did not remove the discontinuity in the derivative
at the cutoff energy, it did reduce the amplitude of residuals. The
centroid energy of this Gaussian was allowed to vary in the fits, but
fell within 0.3\% (usually less) of the cutoff energy.

To model the \crsf, we used a Gaussian shaped function for the optical
depth of the feature. The functional form was:
\begin{equation}\label{eq:gabs}
   \gabseq{}
\end{equation}
where \ecy{} is the energy of the resonance, \dcy{} is the depth
of the line at the resonance, and \wcy{} is the width.  This
profile modifies the underlying continuum in the following way:
\begin{equation}
   \taueq
\end{equation}
We note that, because this is an exponential of a Gaussian, the FWHM
of the line is a complex function of not only the width
\wcy{}, but also the depth \dcy{}. The resulting FWHM is slightly
larger than the normal $2\wcy{}\sqrt{2\ln2}$ for a given \wcy{}, but
asymptotically approaches the standard value as \dcy{} goes to zero.

We also note that although the \mplcut, \fdco, and \npex\ equations
all produce similar curves, there are differences, and a given pulsar
spectrum might be fit well with one and poorly with the others. In
fact, the \mplcut\ fits to 4U~1538$-$52 (\S\ref{subsec:4u1538-52}) and
4U~1907+09 (\S\ref{subsec:4u1907+09}) are somewhat poor. Therefore our
continua fits are not necessarily the best that can be achieved, and
should not be considered the definitive \rxte\ measurements of these
pulsar spectra.

Luckily, the choice of continuum model has little effect on the fits to
the \crsfs\ themselves. This is illustrated in
Table~\ref{table:crsfcont}, where we present the results of fits to 2
different continuum models (the \mplcut\ and \fdco) for a subset of 4
pulsars. Not only are the \crsf\ parameters of each source fit with
the two continua consistent, there is also no systematic trend towards
one continuum giving consistently larger or smaller values of \ecy{}.

The ultimate goal of this analysis was to search for relations among
the various model parameters to reveal real correlations in physical
parameters in the pulsar emission regions. So before starting the
analysis, we examined the systematic properties of these three
modles. Unfortunately none of the models gave consistently smaller
uncertainties for all fit parameters, so any choice would leave some
parameters less well constrained than had we used a different
model. The decision to use the \mplcut\ was based on 3 distinct
advantages it offerend. First, and certainly least important, this
form has been used historically to fit pulsar spectra
\citep[e.g.][]{whi83}. It therefore allows for the direct comparison
of fit parameters of earlier observations, and a look into how things
at the source might have changed \citep[e.g.][]{gru01}.

\begin{figure}[H]
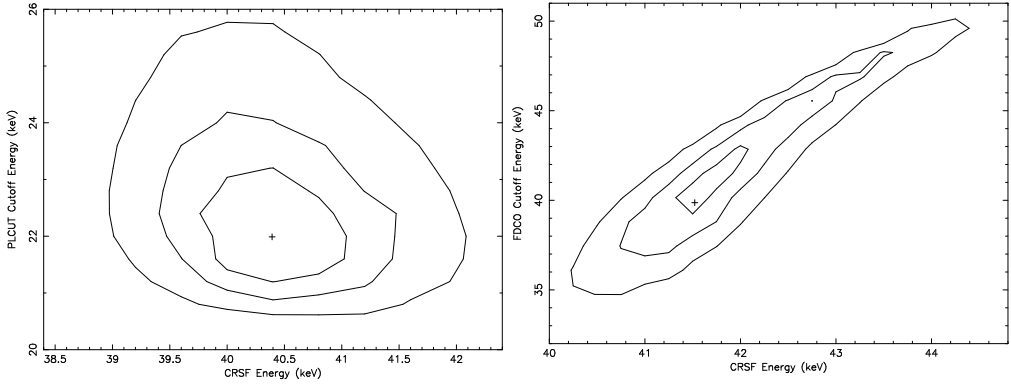

\includegraphics[height=0.40\textwidth,angle=-90]{f2a.eps}
\includegraphics[height=0.40\textwidth,angle=-90]{f2b.eps}
\caption{\label{fig:contplcut}
Left: Contour plot of \mplcut\ cutoff energy \ecut\ versus \crsf\
centroid energy \ecy{}. All plots in Figs.~\ref{fig:contplcut},
\ref{fig:contfdco}, and \ref{fig:contnpex} were obtained using
spectral fits of Hercules X-1, and show 68\%, 90\%, and 99\%
confidence contours. Right: A similar plot using the \fdco\ continuum
model. See \S~\ref{sec:method} for a description of the models. It is
readily apparent the the two parameters are highly correlated with
each other in the \fdco\ model, while for the \mplcut\ model they are
relatively independent. This aspect of the \fdco\ makes attempts to
correlate the values obtained in spectral fitting suspect, hence our
use of a modified \plcut\ (\mplcut) in the analysis presented here.}
\end{figure}

\begin{figure}[H]
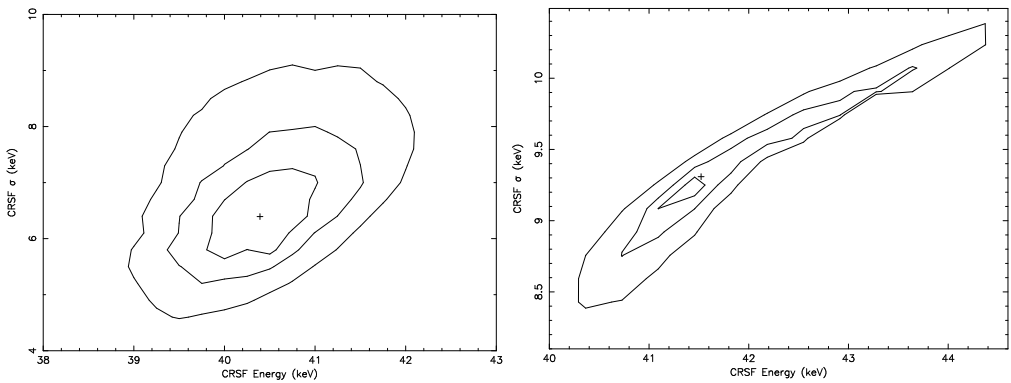

\includegraphics[height=0.40\textwidth,angle=-90]{f3a.eps}
\includegraphics[height=0.40\textwidth,angle=-90]{f3b.eps}
\caption{\label{fig:contfdco}
Left: Contour plot of \crsf\ width \wcy{} versus \crsf\ energy \ecy{}
with a \mplcut\ continuum model (68\%, 90\%, and 99\% confidence
contours). Right: A similar plot using the \fdco\ continuum
model. Again, the parameters of the \fdco\ model exhibit a high degree
of interdependence.}
\end{figure}

\begin{figure}[H]
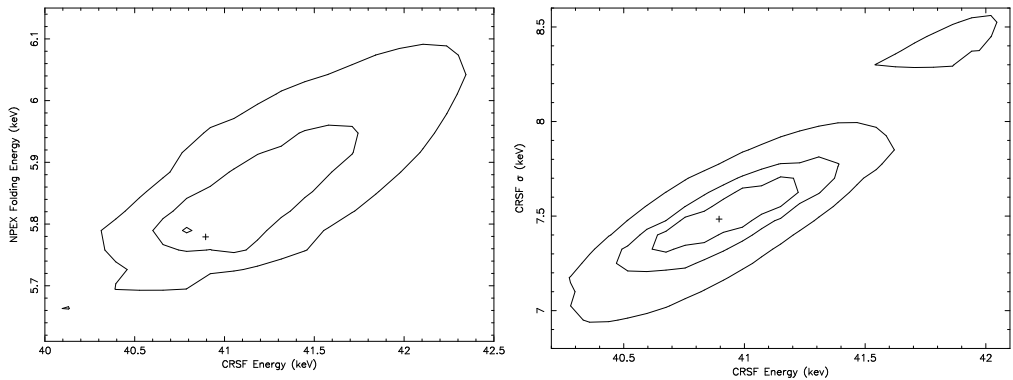

\includegraphics[height=0.40\textwidth,angle=-90]{f4a.eps}
\includegraphics[height=0.40\textwidth,angle=-90]{f4b.eps}
\caption{\label{fig:contnpex}
Similar contours to those above using the \npex\ model, showing the
68\%, 90\%, and 99\% confidence levels. While the contours are not as
pronounced as in the \fdco\ model, they are still quite elongated.}
\end{figure}

Second, the inter-parameter correlations of the other forms meant
that, in general, the \wchi\ valleys for those fits were relatively
broad and flat. Therefore it was difficult for the fitting routines to
find sets of parameters that represented a true minimum in \wchi,
resulting in fits that were slow to converge. These flat error basins
also meant larger uncertainties in the fit parameters, and compared to
the \fdco\ and \npex\ models the \mplcut\ consistently returned smaller
error bars. Comparisons of these error valleys are shown in
Figs.~\ref{fig:contplcut}, \ref{fig:contfdco}, and \ref{fig:contnpex}.

The third and by far the most important consideration was that the
\mplcut\ model parameters themselves are relatively orthogonal to one
another. This can be readily seen in the left hand panels of
Fig.~\ref{fig:contplcut} and \ref{fig:contfdco}, which show error
contours for spectral fits to the \xray\ pulsar Hercules~X-1. In the
case of the \mplcut\ the contours are ``rounder,'' implying a lower
degree of correlation between the model parameters. In contrast, the
contours when fitting with the \fdco\ model are quite elongated, with
the axes of the ellipse at an angle to the parameter axes (see
Fig.~\ref{fig:contplcut} and \ref{fig:contfdco}, right panels). The
\npex\ model is slightly better than the \fdco, but it is still not
ideal for this type of analysis (see Fig.~\ref{fig:contnpex}).

The problem arises when analyzing an ensemble of fits using models
that exhibit such systematic correlations. Even if the underlying
physical parameters are random, inferred parameters of poorly chosen
models can show strongly correlated results. These correlations,
however, are merely artifacts of the modeling procedure and \emph{not}
intrinsic to the data set. This also means that any real signal in the
data will be suspect because of the potential for spurious results. By
using a \mplcut\ continuum, we have removed much of the
systematic effects from the fitting process. Therefore the work
presented here best reflects the underlying nature and physics of the
pulsar polar emission region, and not artifacts of the fitting process
itself.

%% /*******************************************************************
%% ** Observations                                                   **
%% *******************************************************************/

\section{Sources and Observations}\label{sec:obs}

In this section we describe our sample of 10 accreting pulsars,
previous observations of \crsfs, the observations made using the
\rxte, and the observed spectra. These pulsars have received various
amounts of coverage by the \rxte\ in the first 5 years of
operation. Given the size and richness of this database (over 4.2\,Ms
of satellite time), we instead discuss a representative subset of
observations (see Table~\ref{table:obs}), chosen to be as long and
continuous as possible, and with each source as bright as
possible. The results of our spectral parameterizations are given in
Table~\ref{table:plcutfits}, and the remaining fit parameters (\nh,
\kte, etc.) are listed in Table~\ref{table:params}.

%%
%% Hercules X-1
%%
\subsection{Hercules~X-1}\label{subsec:herx1}

Her~X-1 is characterized by a 1.24\,s pulse period \citep{tan72} and
an eclipsing 1.7\,d orbit \citep{dee81} with the A/F star HZ Her
\citep{dox73,got91}. In addition to the \xray\ eclipses, Her~X-1 also
exhibits a 35\,d intensity cycle due to the
precession of a tilted, warped accretion disk viewed nearly edge on
and which periodically obscures \xrays\ from the central neutron star
\citep{pet75}. Recent models of the physical cause of the warping have
been suggested by, for example, \citet{sha99}, \citet{mal97},
\citet{pri96}, and \citet{sch94}. Due to this disk precession, the
source exhibits a \wsim10\,d ``main-on'' (where \xray\ emission is a
maximum) and a dimmer \wsim5\,d ``short-on'', separated by two
\wsim10\,d ``low'' states (where virtually no \xray\ emission is
detected). Her~X-1 was the first \xray\ pulsar in which a \crsf\ was
discovered \citep{tru78}, and since then this feature has been well
studied. A recent analysis by \citet{gru01}, comparing \rxte\
observations with those of previous missions, has shown that the
cyclotron resonance energy changed from \wsim34\,keV to \wsim42\,keV
sometime between 1991 and 1993.

For the analysis here, only data taken during the peak of a main-on
and out of eclipse were used. This was done to ensure that the
observed spectrum represented the intrinsic neutron star polar cap
emission, minimizing the effect of the accretion disk. The spectrum,
along with the parameterizing fit, is shown in
Fig.~\ref{fig:specfigsone}, upper left panel. In addition to a \mplcut\
continuum and \crsf, the fit contains two Fe-K lines (both near the
iron fluorescence energy 6.5\,keV). The first line is narrow, with a
78\,eV equivalent width (EW), and the second one much broader and with
a larger (370\,eV) EW. The broad Fe-K line is similar to the single
240\,eV EW Fe-K line seen with \ginga\ \citep{mih91} during the
main-on, although we find that a second, narrow line (which is at the
detection limit for an Fe-K line in the \pca) is required by the
\rxte\ data to get an acceptable \wchi. Overall the fit is quite good,
and we find a \crsf\ at $\ecy{}=40.4_{-0.3}^{+0.8}$\,keV. This agrees
with the analysis of \citet{gru01} for the same dataset. There is no
indication of a second \crsf\ harmonic near 80\,keV, although due to
the steeply falling spectrum our sensitivity to such a feature is
limited. A preliminary analysis of a \sax\ observation does, however,
indicate that a harmonic might be present in the spectrum (Santangelo,
private communication). There is also a small pattern in the residuals
near 10\,keV. This feature is only a \wpm0.8\% deviation, and so while
it does affect the resulting \wchi, the fit parameters are not
significantly affected. It does, however, appear systematically
through \wsim10 PHA channels. This unexplained feature near 10\,keV is
exhibited in the spectra of many accreting
\xray\ pulsars, and is discussed in \S\ref{subsec:10keV}.

%%
%% 4U0115+63
%%
\subsection{4U~0115+63}\label{subsec:4u0115+63}

The transient \xray\ source 4U~0115+63 is an accreting \xray\ pulsar
with a 3.6\,s pulsation period in an eccentric 24\,d orbit
\citep{bil97} with the O9e star, V635 Cassiopeia
\citep{ung98}. \xray\ outbursts have been observed from 4U~0115+63
with \uhuru\ \citep{for76}, \heao--1 \citep{whe79,ros79},
\ginga\ \citep[e.g.][]{tam92}, and \cgro/\batse\ \citep{bil97}, and
reoccur with a separation times of one to several years. The \rxte\
has observed two outbursts of 4U~0115+63, the first in 1999 March
\citep{wil99iauc,hei99iauc}, and the second in 2000 September
\citep{cob00iauc}.

A \crsf\ in 4U~0115+63 was discovered near 20\,keV by
\citet{whe79} with the UCSD/MIT \heao--1/A4 low-energy (13--180\,keV)
detectors. \citet{whi83} analyzed concurrent data from the lower
energy (2--50\,keV) \heao--1/A2 experiment and found an additional
feature at \wsim12\,keV, making this the first pulsar to exhibit more
than a single line. We discuss here observations of the 1999 March
outburst obtained with the \rxte. It was during this outburst that
4U~0115+63 became the first \xray\ pulsar found to exhibit more than
just a fundamental and single overtone \crsf\
\citep{hei99,san99}, with 5 features observed before the spectrum runs
out of statistics \citep{hei99comp0115}.

The fit shown in Fig.~\ref{fig:specfigsone} (upper right) consists of
a \mplcut\ continuum along with 4 \crsfs. This dataset has been
discussed previously in \citet{hei99} and \citet{hei99comp0115}, and
although different continuum models were used our results here are
similar. Here, using a simple line shape (see Eq.~\ref{eq:gabs})
resulted in a \crsf\ energies of $16.4^{+0.4}_{-1.0}$\,keV,
$23.2_{-0.9}^{+0.4}$\,keV, $31.9_{-0.7}^{+0.7}$\,keV, and
$48.4_{-1.3}^{+0.8}$\,keV. The three higher harmonics (at 23, 32, and
48\,keV) are roughly consistent with the hypothesis of harmonic
spacing, although the fit fundamental is at too high of an
energy. Analyzing the source spectrum as a function of neutron star
rotation phase, \citet{hei99comp0115} found that the actual line shape
is complex and poorly modeled by a Gaussian shaped absorption
feature. Additionally, the authors found that the line centroids
change considerably with pulse phase and are therefore difficult to
model exactly using phase averaged spectroscopy.

So, while a single broad line centered at 16\,keV is a good
representation of the line shape (which is of primary interest here),
the cyclotron energy of the magnetic field is better described as half
that of the second harmonic, or $\ecy{}=11.6_{-0.4}^{+0.2}$\,keV. In
the fit there is also a pattern in the residuals near 7\,keV that
resembles an Fe-K line. These are very small deviations, however, and
an Fe-K line is not formally required by the fit (with an $F$-test
probability of $5\times10^{-3}$). We note that fits done by
\citet{hei99} and \citet[][(using \sax\ observations obtained during
the same outburst]{san99} do not require the addition of an Fe-K line.

%%
%% Cen X-3
%%
\subsection{Cen~X-3}\label{subsec:cenx3}

The eclipsing source Cen~X-3 has a spin period of 4.8\,s and a 2.1\,d
orbital period \citep{kel83cen} around the O6-8 supergiant V779 Cen
\citep{krz74,ric74,hut79}. Historically the 1--40\,keV spectrum has
been fitted with the standard pulsar continuum, Fe-K line, and low
energy absorption \citep{whi83}. The analysis of \ginga\ data by
\citet{nag92} showed that the high-energy cutoff is better modeled by
a broad, pseudo-Lorentzian turnover rather than by the standard
e-folding energy cutoff. They did not claim a firm detection of a
\crsf\ since the energy of the resonance ($\gtrsim30$\,keV) is near
the 37\,keV limit of the \ginga\ LAC detectors. A \crsf\ at 30\,keV
was discovered by \citet{san98} and \citet{biff99} using \sax\ and
\rxte\ observations.

In addition to a \mplcut\ continuum and \crsf, we included an Fe-K line
and low-energy absorption in the fit (see Fig.~\ref{fig:specfigsone},
middle left). The equivalent width of the Fe-K line was 141\,eV, which
is similar to the value obtained by \citet{bur00} using observations
obtained with \sax. The amount of low-energy absorption required,
$\nh=(2.2\pm0.2)\times10^{22}$\,\pcmsq, is consistent with the
\sax\ value of $(1.95\pm0.03)\times10^{22}$\,\pcmsq. As with
Her~X-1 we find a small yet systematic pattern in the residuals
near 10\,keV, although for Cen~X-3 the ``peak'' is at a slightly
higher energy.

The pulse phase average \crsf\ energy is $\ecy{}=30.4_{-0.4}^{+0.3}$,
and is consistent with what was found with \sax\ \citep{san98}. \rxte\
and \sax\ pulse phase resolved analyses have both revealed that the
energy moves through the pulse; from $29.0_{-0.3}^{+0.4}$\,keV during
the pulse fall to $37.1\pm1.7$\,keV during pulse rise
\citep{biff99}. This type of variability is not unique to this source
(other examples are Her~X-1, \citealt{soo90a}, and 4U~1538$-$52,
\citealt{cla90}), although the magnitude of the variation in Cen~X-3
(\wsim20\%) is quite large. This variability of the \crsf\ centroid
energy implies that, as the neutrons star rotates, different regions
of the accretion structure are observed and different magnetic field
strengths are indicated in these regions. This may be evidence for the
dipole moment being significantly offset from the center of the
neutron star \citep{bur00}. It might be expected that this variation
in the \crsf\ energy could artificially broaden a phase average
measurement of the \crsf\ line width. The average emission is,
however, dominated by flux from the peak of the pulse. A comparison of
our fits with those of \citet{biff99} and \citet{bur00} indicate that
this movement did not significantly broaden the parameterization of
the \crsf\ width \wcy{}.

%%
%% 4U 1626-67
%%
\subsection{4U~1626$-$67}\label{subsec:4u1626-67}

The 7.66\,s pulsar 4U~1626$-$67 is in a binary system with the
low-mass dwarf star KZ~TrA. Although a complete orbital ephemeris
is unknown there is evidence for a 42\,min orbital period
\citep{mid81}. The system has one of the lowest known mass
functions for a pulsar \citep{lev88}, with a limit of $a_{\rm
x}\sin(i)<8$\,lt-ms \citep{shi90}.  Optical limits on the companion's
luminosity rule out the possibility of accretion by a normal stellar
wind \citep{cha98}, so mass flow is most probably due to Roche lobe
overflow of the companion in a very close binary system.

\begin{figure}[H]
\includegraphics[width=0.47\textwidth]{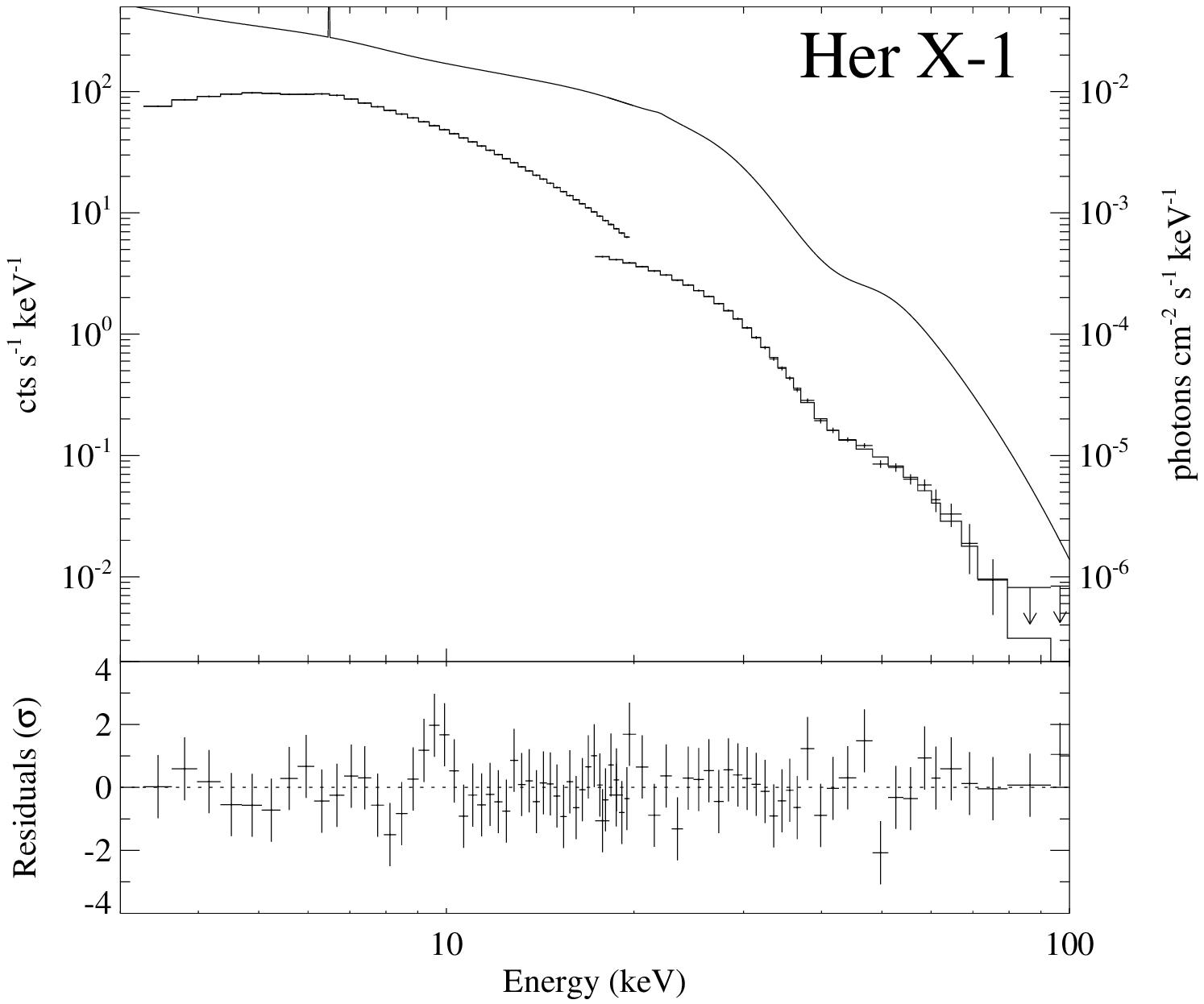}
\includegraphics[width=0.47\textwidth]{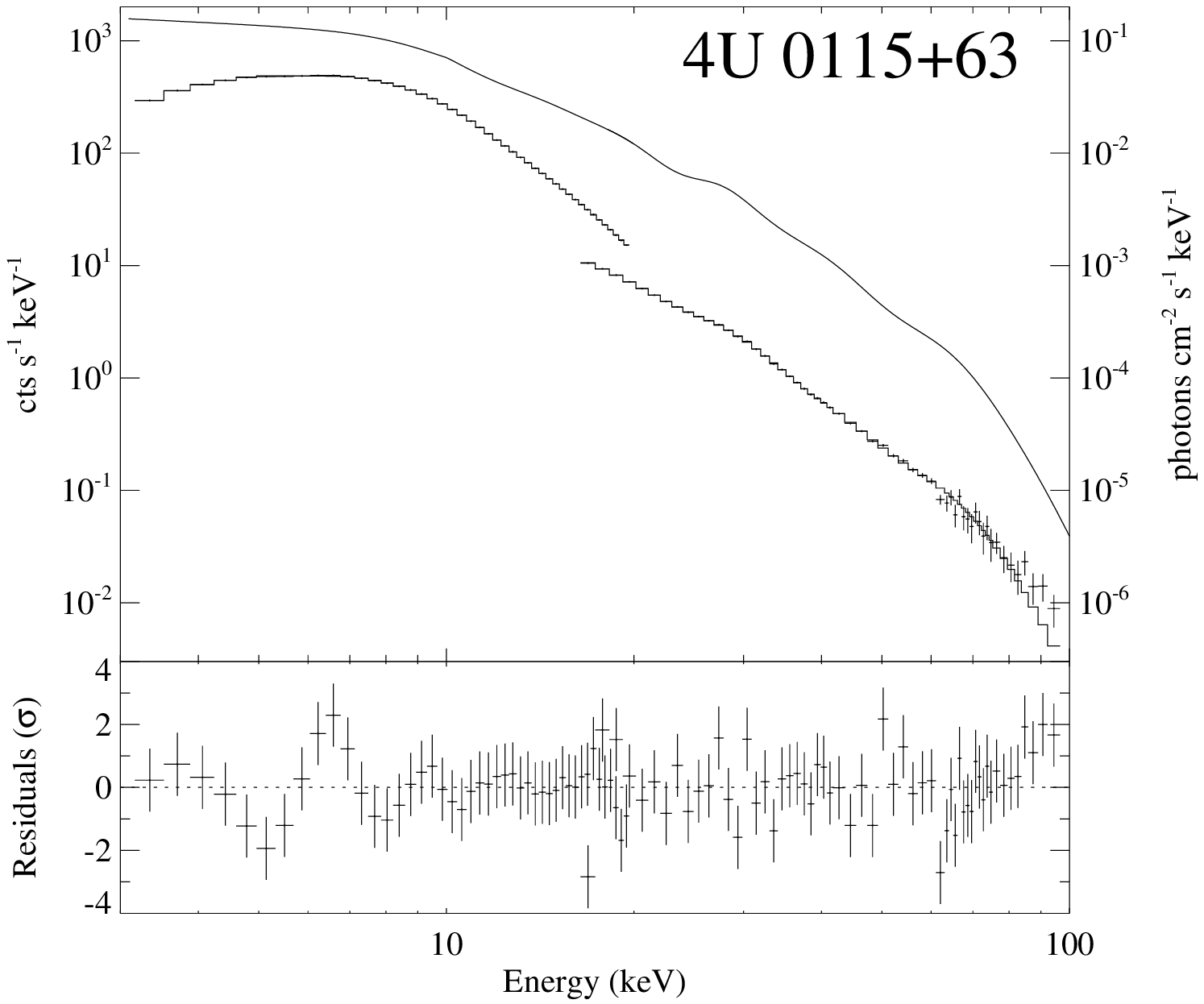}\\
\includegraphics[width=0.47\textwidth]{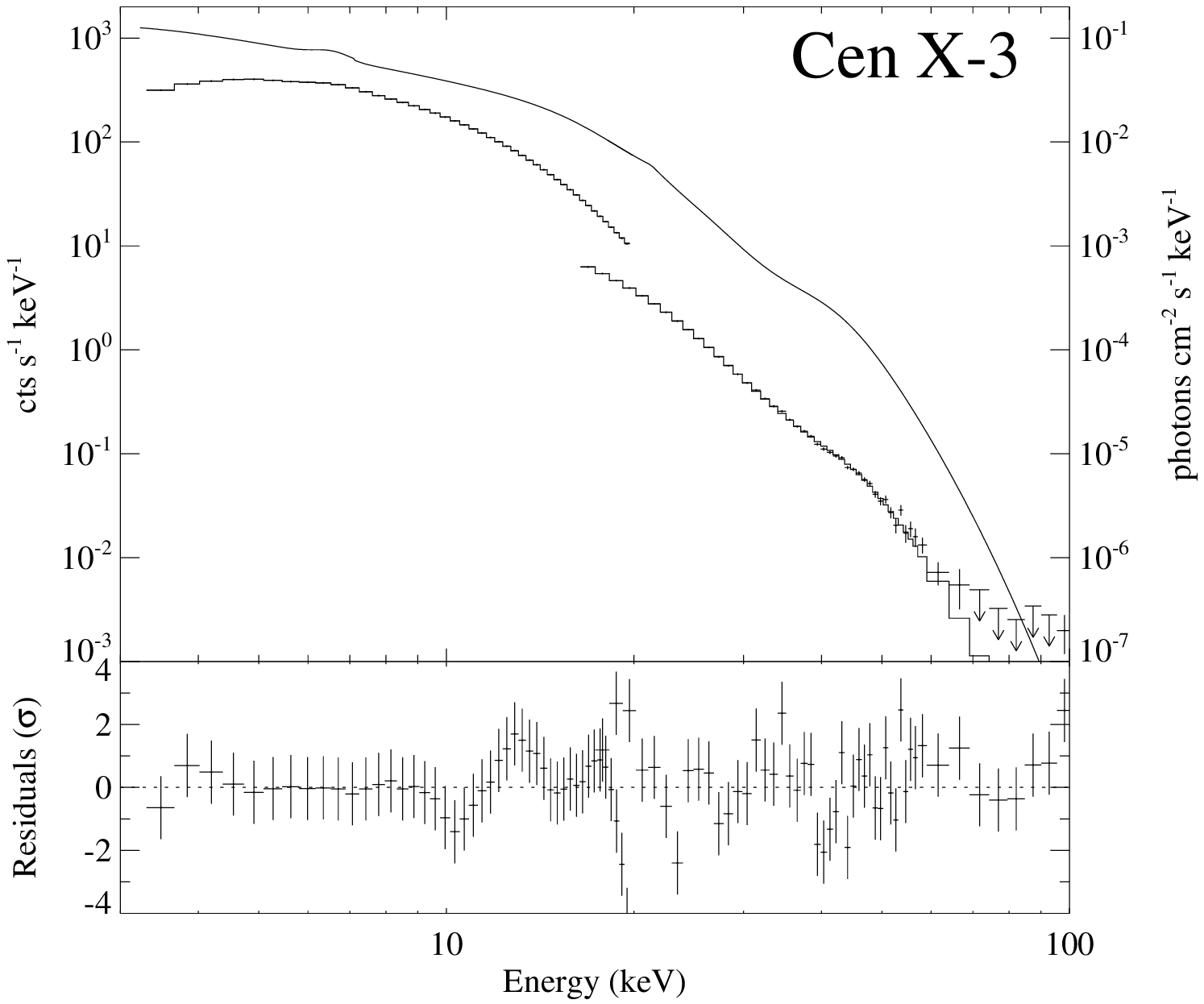}
\includegraphics[width=0.47\textwidth]{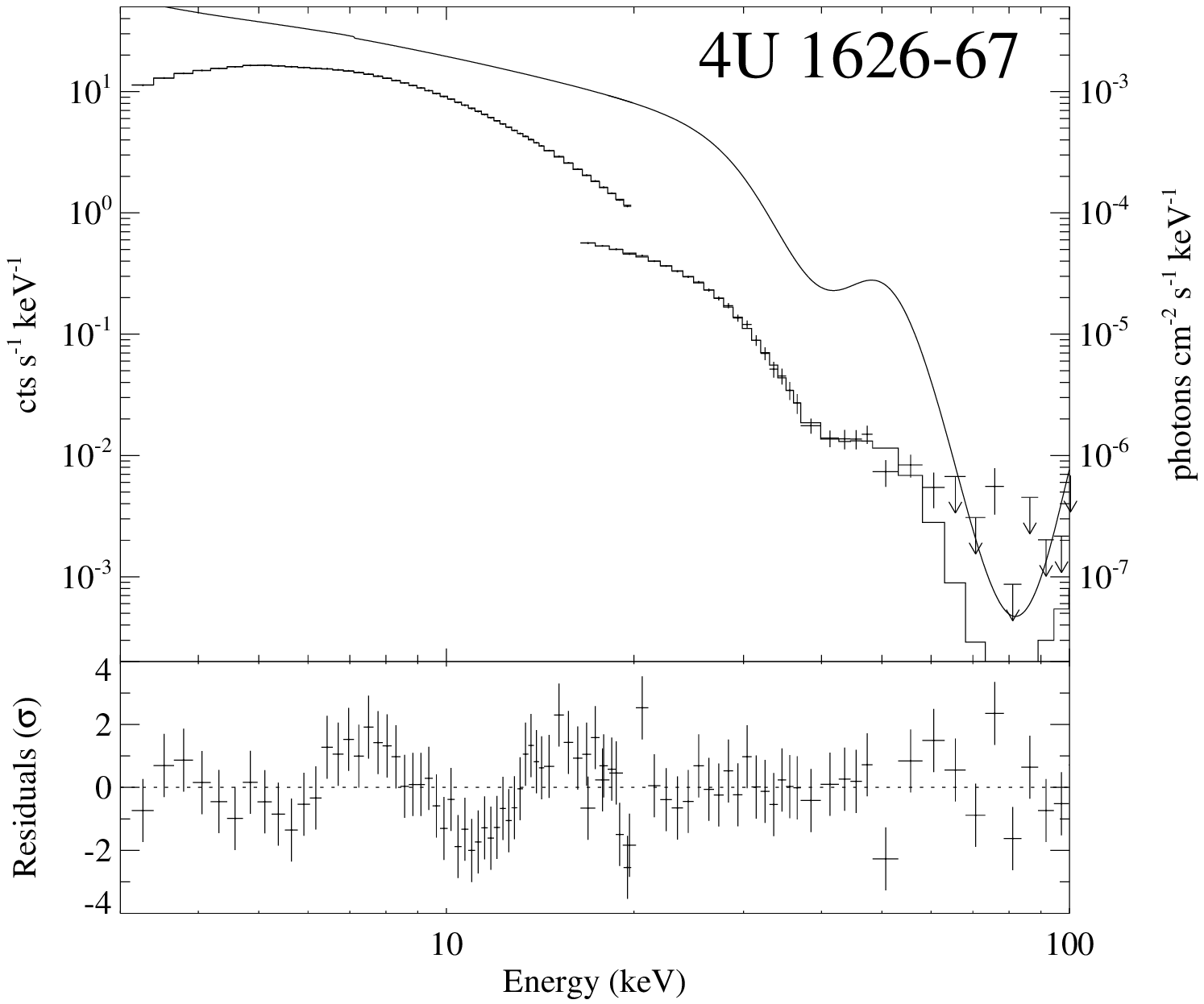}\\
\includegraphics[width=0.47\textwidth]{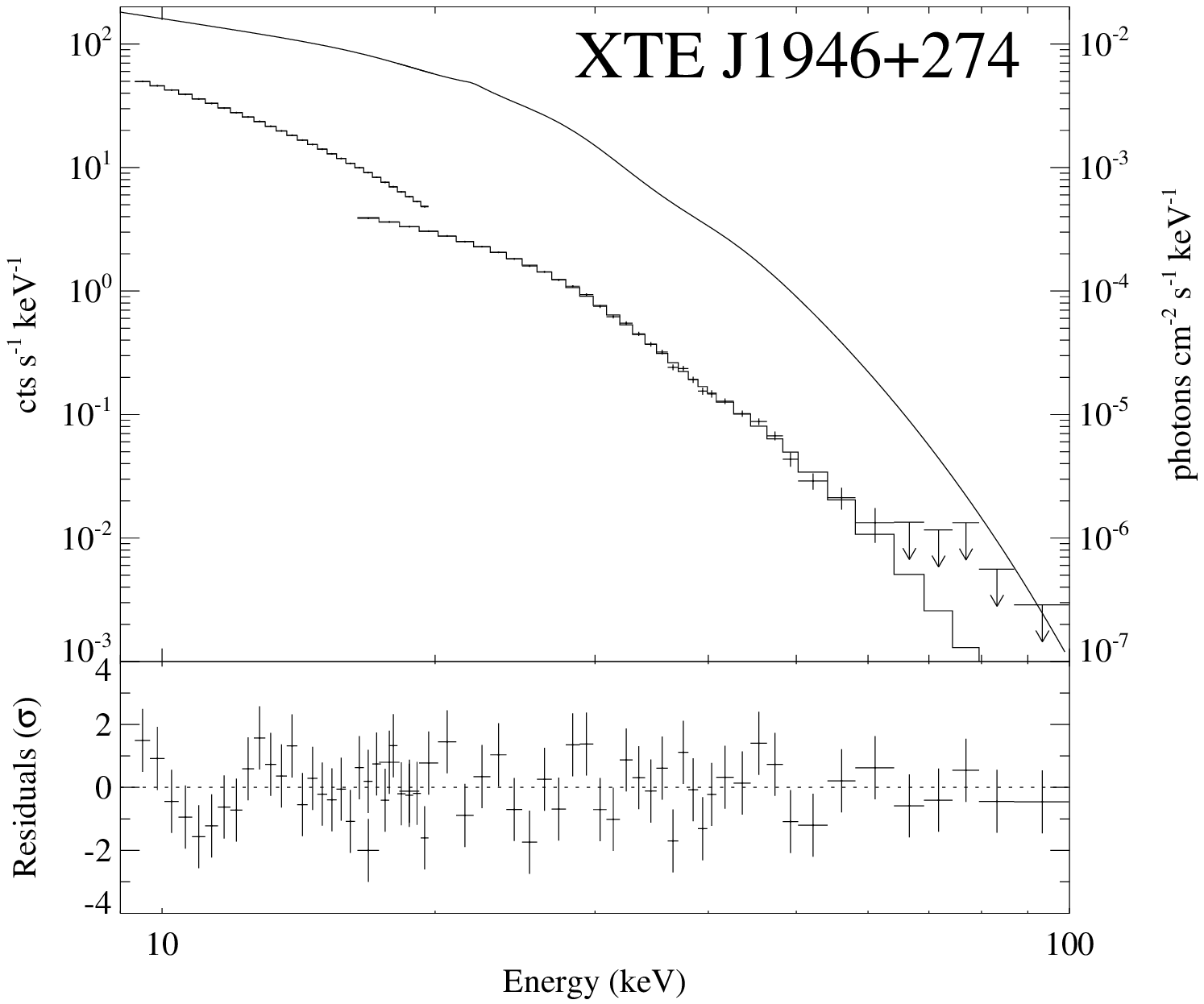}
\includegraphics[width=0.47\textwidth]{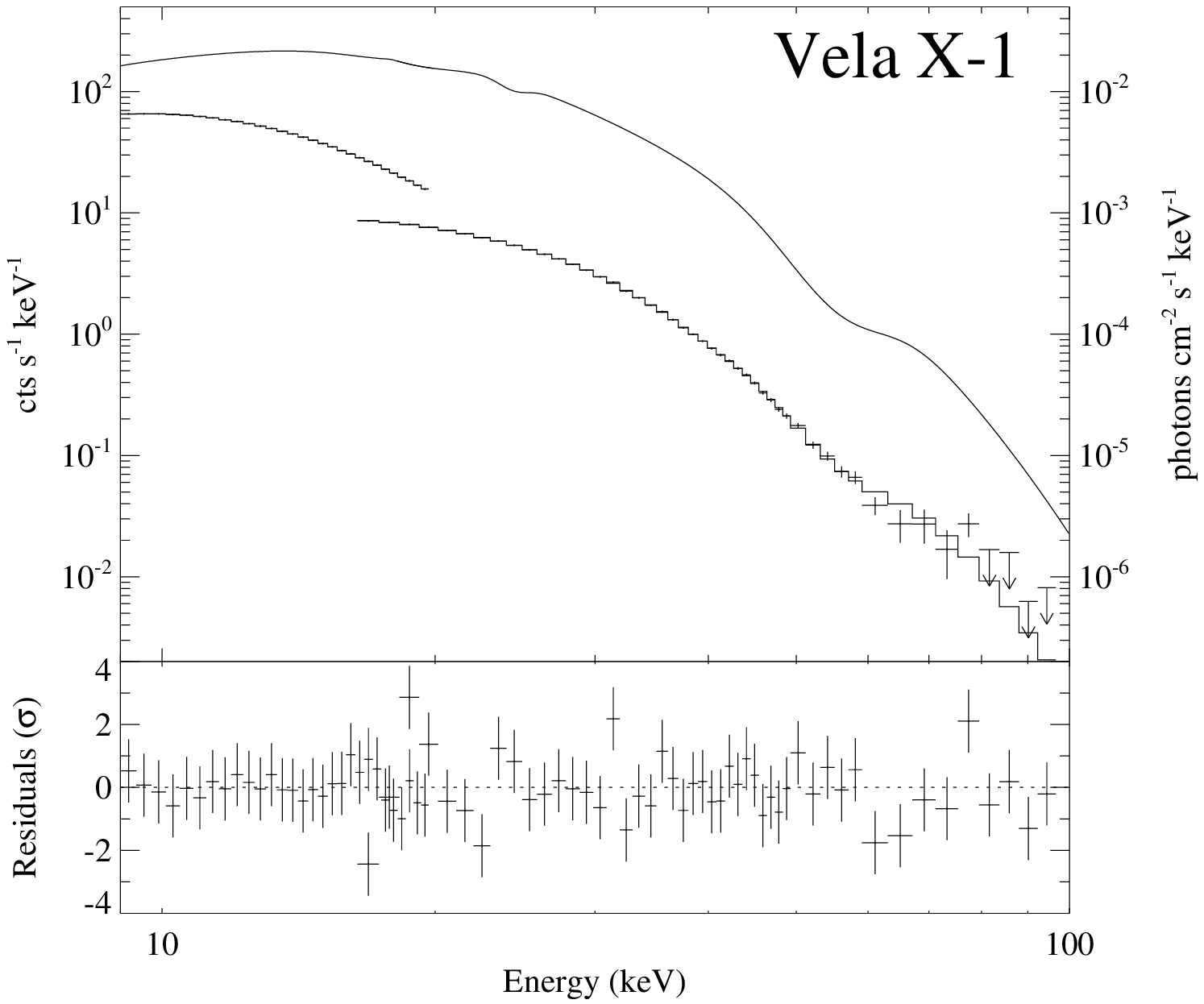}
\caption{\label{fig:specfigsone}
Upper Left: Her~X-1. Upper Right: 4U~0115+63. Middle Left: Cen~X-3.
Middle Right: 4U~1626$-$67. Lower Left: XTE~J1946+274. Lower Right:
Vela~X-1. All Panels: The top pane presents the counts spectrum
(crosses), inferred photon spectrum (smooth curve, see appropriate
source section for details) and model folded through the instrumental
responses (histogram). The bottom pane contains the residuals to the
fit in units of $\sigma$.}
\end{figure}

The spectrum presented in Fig.~\ref{fig:specfigsone} (middle right
panel) was obtained early in the \rxte\ mission, during \pca\ gain
epoch 1. This was also before the failure of one of the pulse height
analyzers in cluster B of \hexte\ (see \S\ref{sec:rxte}), so the
\hexte\ spectrum consists of spectral information from all 8 phoswich
detectors.

Historically the hard \xray\ spectrum of 4U~1626$-$67 has been modeled
with a \wsim0.6\,keV blackbody and low energy absorption in addition
to the standard pulsar continuum \citep[e.g.][]{pra79,kii86}, so we
allowed for these components as well. Our best fit blackbody
temperature was $kT=0.35_{-0.05}^{+0.04}$\,keV. This is consistent
with the \sax\ value of $0.33\wpm0.02$\,keV \citep{owe97}. The
measured absorption column was
$(1.6_{-0.7}^{+0.9})\times10^{22}$\,\pcmsq, which is consistent with
the $8\times10^{21}$\,\pcmsq\ column measured by \sax\
\citep{owe97}. As seen previously, we find a \crsf\ at $\sim39$\,keV
in the spectrum \citep{biff99,orl98}. In addition to the fundamental,
we also found that the addition of a \emph{second} harmonic
significantly improves the continuum fit, with an \ftest\ probabilty
of $2\times10^{-13}$. While this \crsf\ at $80_{-4}^{+16}$\,keV is
statistically significant, it is also only evidenced by the effects of
its red wing on the continuum (see Fig.~\ref{fig:specfigsone}, middle
right panel) and so its existence is still somewhat questionable.

%%
%% XTE J1946+274
%%
\subsection{XTE~J1946+274}\label{subsec:xtej1946+274}

Not much is known about the transient pulsar XTE~J1946+274. The source
was discovered with the \rxte/\asm\ during a scan of the Vul-Cyg
region on 1998 September 5. It was detected at a 2--12\,keV flux level
of \wsim13\,mCrab \citep{smi98}. It reached a peak of \wsim110\,mCrab
around 17 September 1998 \citep{tak98iauc}. Pulsations at 15.8\,s were
discovered by the \cgro/\batse\ \citep{wil98} and were later confirmed
through pointed \rxte\ observations \citep{smi98}. There is also
evidence for a \wsim80\,d periodicity in the \asm\ lightcurve, which
is most likely due to the orbital motion of the binary system
\citep{cam99}. A $36.2_{-0.7}^{+0.5}$\,keV \crsf\ has been detected in
the \xray\ spectrum with both the \rxte\ \citep{coburn01,hei01xtej}
and \sax\ \citep{san01}.

The spectrum presented here (see Fig.~\ref{fig:specfigsone}, lower
left) was obtained by summing 11 individual observations obtained over
the course of about one month. \citet{hei01xtej} found that changes in
the spectrum from observation to observation were consistent with
changes in \nh\ of order 10\%. Such variations are not unexpected
during a Be/\xray\ binary outburst, and they are likely not indicative
of changes in the underlying continuum emission. For this reason, we
concluded that, by excluding data below 8\,keV, we could confidently
combine the data, and fit with a \mplcut\ continuum and
\crsf.

%%
%% Vela X-1
%%
\subsection{Vela~X-1}\label{subsec:velax1}

Vela~X-1 (4U~0900$-$40) is an eclipsing \xray\ binary with an orbital
period of 8.964\,d \citep{ker95} and which exhibits a pulsation period
of \wsim283\,s. The companion star is the B0.5Ib supergiant HD~77581,
and the system is at a distance of 2.0\,kpc \citep{sad85}. Due to the
closeness of the binary orbit, the neutron star is deeply embedded in
the strong stellar wind of its companion
\citep[$\dot{M}\sim4\times10^{-6}\,$\msol\,yr$^{-1}$,][]{nag86}.
\xrays\ are produced by accretion of this wind onto the
neutron star surface. The \xray\ luminosity is typically $L_{\rm
x}\sim4\times10^{36}$\,\lum, but can also suddenly change to less than
10\% of its normal value \citep{ino84}. The cause of this variability
is still unknown, although it could be due to changes in the accretion
rate due to variations in the stellar wind, or possibly associated
with the formation of a temporary accretion disk \citep{ino84}.

The \xray\ spectrum of Vela~X-1 is usually described by the standard
accreting pulsar continuum plus an Fe-K line
\citep{nag86,whi83,oha84}. Near periastron the spectrum is modified by
significantly increased absorption due to a gas stream trailing the
neutron star as well as circumstellar matter \citep{kap94}. In
addition, erratic increases in \nh\ by a factor of 10 are seen at all
orbital phases \citep{hab90}. Evidence for a \crsf\ at \wsim54\,keV
has been reported from the \HEXE\ (\hexe) on the Mir space station
\citep{ken92}, from \ginga\ \citep{mih95}, from the \rxte\
\citep{kre99,kre01}, and from \sax\ \citep{orl98vela}. There was
further evidence that the fundamental \crsf\ lies at \wsim24\,keV
\citep[e.g.][]{ken92,mih95}, with the line at 54 being the second
harmonic. Using a pulse phase resolved analysis, \citet{kre01} have
now confirmed the existence of the lower energy line.

The \rxte\ data we used here has already been discussed in detail by
\citet{kre01}. The continuum we use here is different, however our
fits to the \crsfs\ are consistent with the previous results. In the
first half of the observation an anomalous \xray\ dip was observed,
lasting \wsim17\,ks, and during this time no pulsations were seen
\citep{kre00comp}. The spectrum we present here (see
Fig.~\ref{fig:specfigsone}, lower right) was accumulated after the
dip, where the average counting rate was steady and pulsations were
observed. We parameterized the phase average spectrum of Vela~X-1 with
a \mplcut\ continuum and two \crsfs. To minimize the effects on the fit
continuum spectrum due to the highly variable absorption column and
Fe-K line, we ignored \pca\ data below 9\,keV. However, we find that
some \xray\ absorption is still required by the data, with
$\nh=(2.9_{-0.2}^{+0.3})\times10^{23}$\,\pcmsq. The width
(\wcy{}=$0.9_{-0.8}^{+0.9}$\,keV) and depth
(\dcy{}=$0.16_{-0.07}^{+1.13}$) of the fundamental \crsf\ are poorly
constrained, although the addition of an absorption line near 24\,keV
gives an $F$-test probability of $1.14\times10^{-7}$ for a chance
improvement.

%%
%% 4U 1907+09
%%
\subsection{4U~1907+09}\label{subsec:4u1907+09}

The \xray\ source 4U~1907+09 was discovered with the \uhuru\ satellite
\citep{gia71}, while observations with \tenma\ revealed the source to
be a pulsar with a 437.5\,s period \citep{mak84}. Recent \rxte\
observations have produced an improved binary orbital ephemeris
\citep{zan98}, along with the discovery of \xray\ dips \citep{zan97}.
The neutron star is powered by wind accretion from either a close OB
supergiant \citep{sch80} or possibly a Be star
\citep{iye86,rob01astroph}, and is in a close orbit (8.4\,d) with the
companion star. The light curve for our dataset is free from \xray\
flares or dips, so the spectrum is not affected by possible spectral
variability due to intensity variations. A \crsf\ at \wsim21\,keV was
discovered in this source by \citet{mak92} using \ginga\ observations,
while \citet{cus98} detected both the fundamental and a second
harmonic at \wsim39\,keV. We find a \crsf\ energy of
$18.3\pm0.4$\,keV.

\begin{figure}[H]
\includegraphics[width=0.47\textwidth]{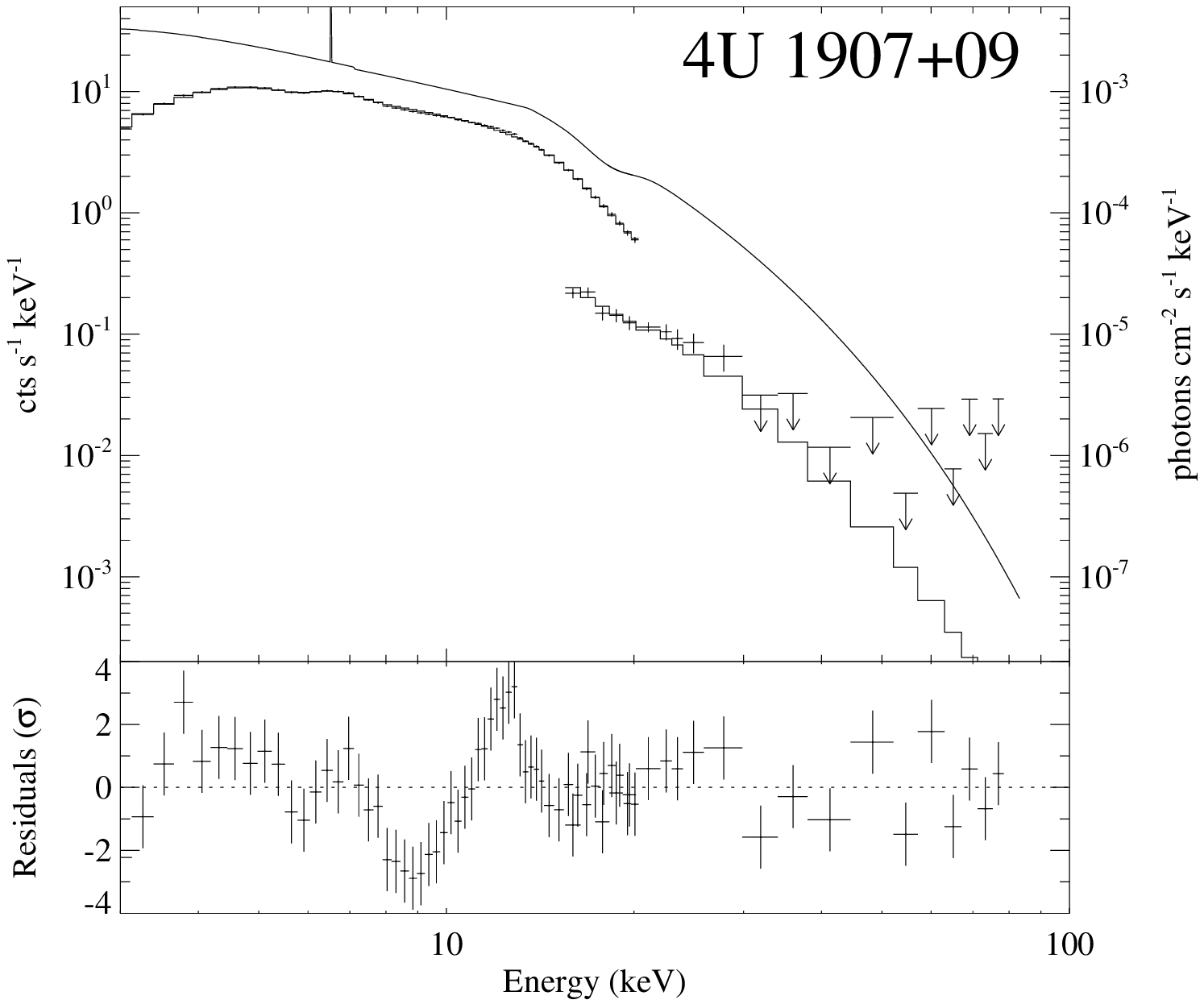}
\includegraphics[width=0.47\textwidth]{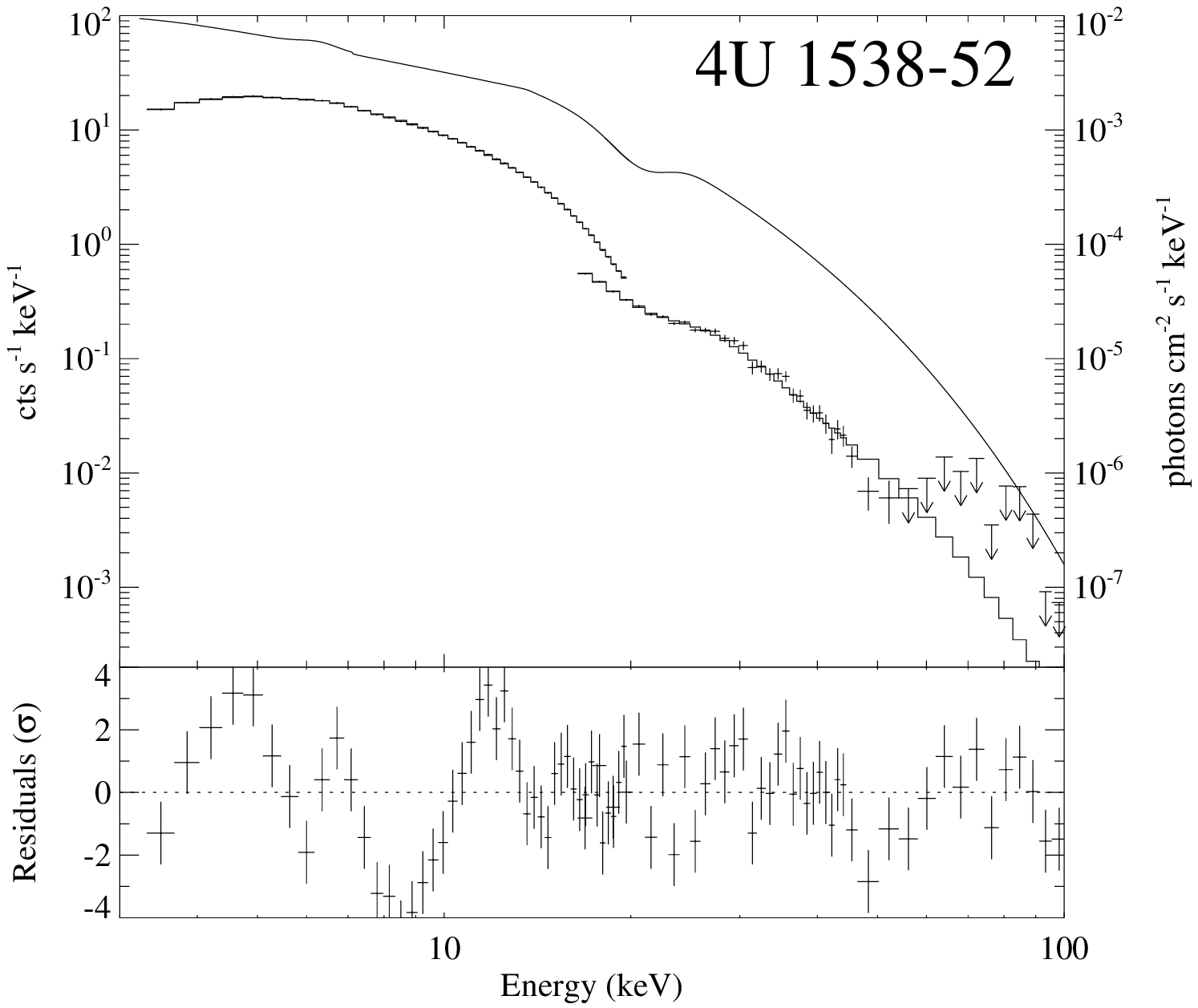}\\
\includegraphics[width=0.47\textwidth]{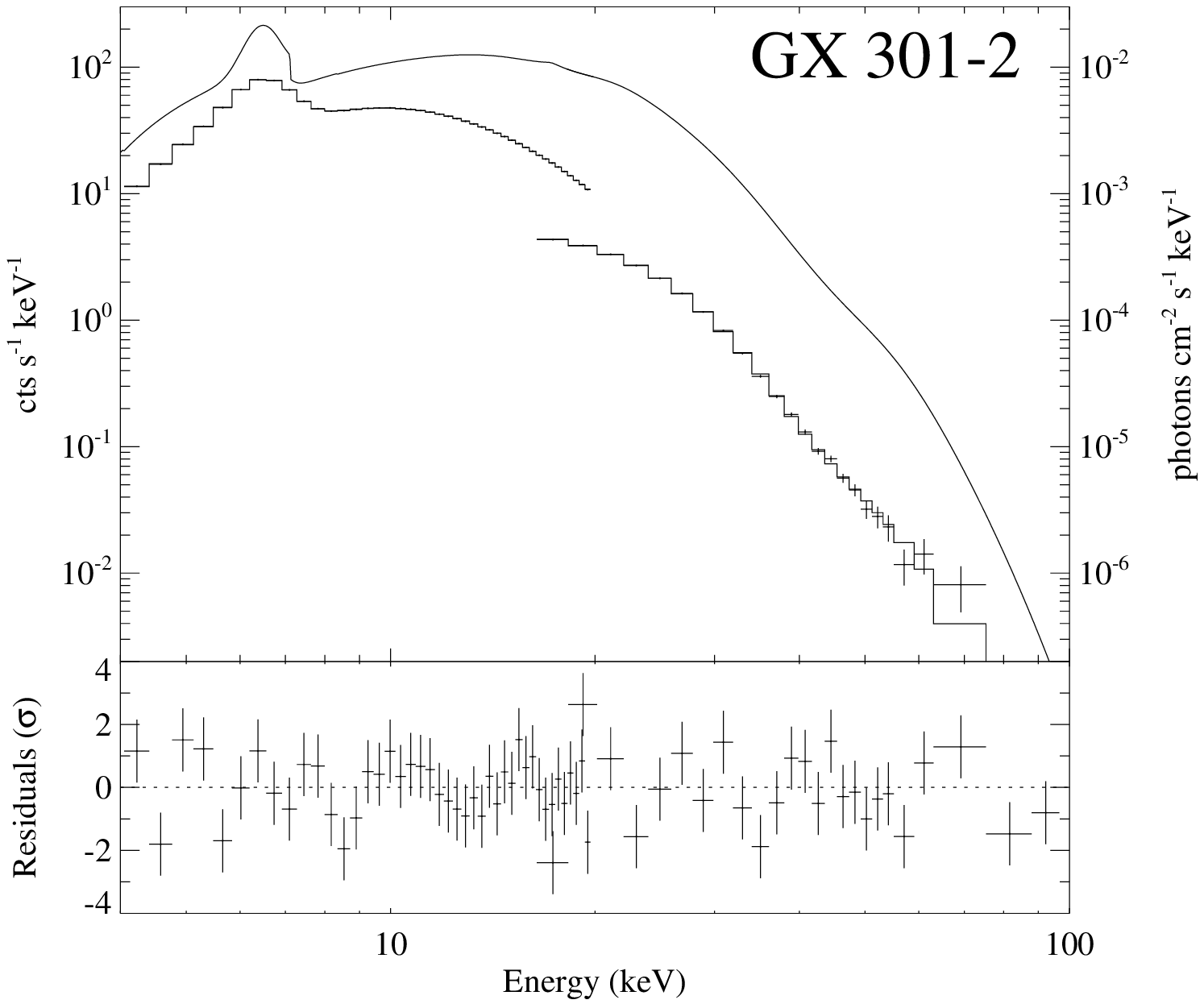}
\includegraphics[width=0.47\textwidth]{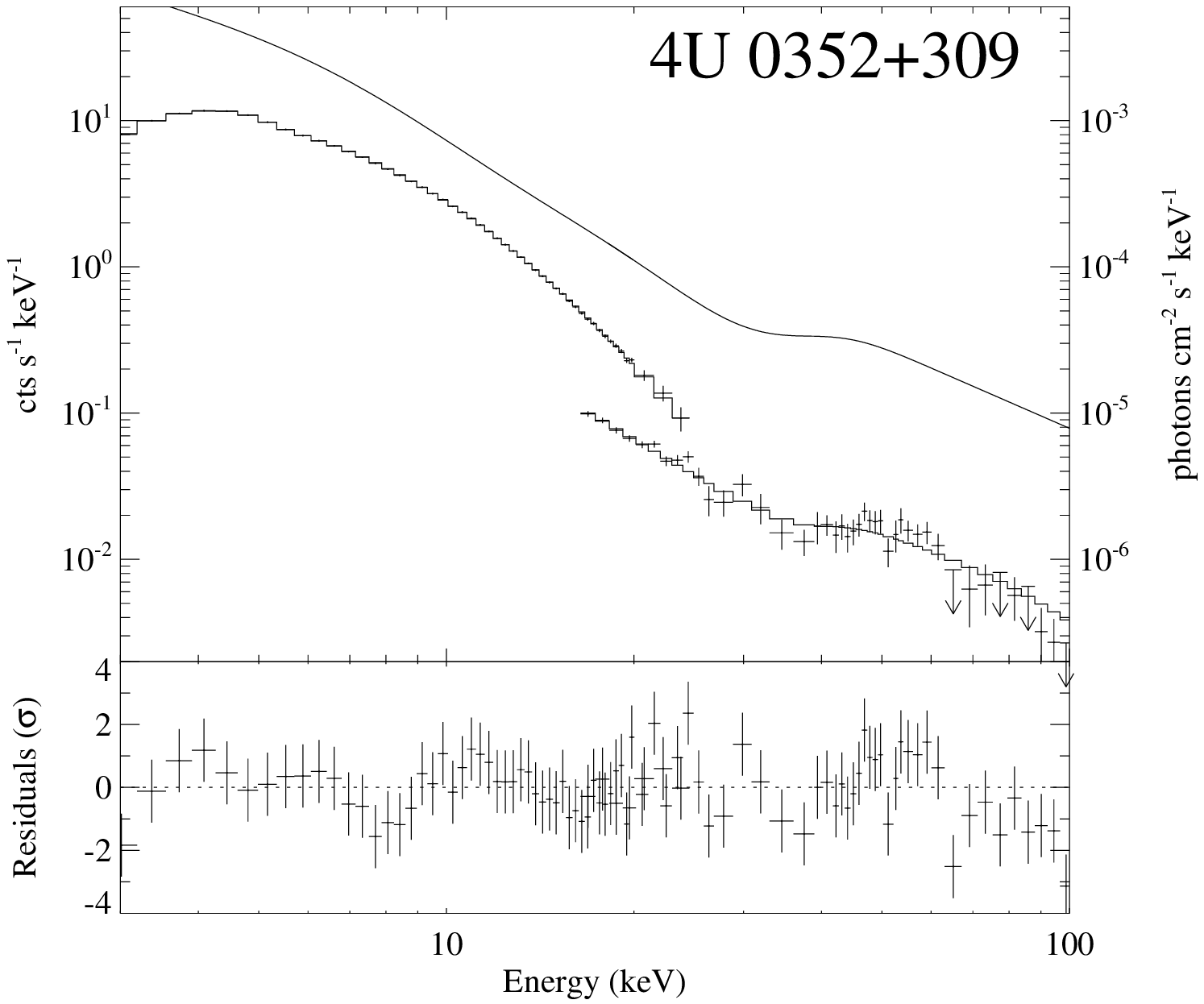}\\
\caption{\label{fig:specfigstwo} 
Upper Left:4U~1907+09. Upper Right: 4U~1538$-$52. Lower Left:
GX~301$-$2.  Lower Right:4U~0352+309. All Panels: The top pane
presents the counts spectrum (crosses), inferred photon spectrum
(smooth curve, see appropriate source section for details) and model
folded through the instrumental responses (histogram). The bottom pane
contains the residuals to the fit in units of $\sigma$. }
\end{figure}

Like 4U~1626$-$67 (\S\ref{subsec:4u1626-67}), the spectrum here (see
Fig.~\ref{fig:specfigstwo}, upper left) was obtained before the
failure of a pulse height analyzer in one of the \hexte\ detectors
(see \S\ref{sec:rxte}), so spectral information from all 8 phoswich
detectors is present. In addition to the standard pulsar continuum and
\crsf, there is evidence for soft \xray\ absorption in the range
$\nh=(1.5-1.7)\times10^{22}$\,\pcmsq\
\citep[see][]{sch80,mar80,mak84,cook87,chi93} and an Fe-K emission
line. We find a column of $(2.4\pm0.1)\times10^{22}$\,\pcmsq, which is
larger than the historical values. We also find a narrow Fe-K line
improves the fits. The residuals to this parameterization below
10\,keV, however, make the interpretation of spectral features in this
energy range ambiguous (see Fig.~\ref{fig:specfigstwo}, upper left).

There is a large (\wpm3\%) pattern in these residuals that ``peaks''
slightly above 10\,keV, making this the most pronounced case of the
10\,keV feature. It is tempting to associate this with a \crsf\ at
\wsim9\,keV, making the features at 18\,keV and 39\,keV the second and
fourth harmonics respectively (with the third harmonic absent). When
we fit a line in this region, the energy centroid energy was indeed
\ecy=$8_{-4}^{+2.}$\,keV. The feature is, however, quite broad
(\wcy=$7\pm3$\,keV), and rather than appearing as a simple line it
modifies the continuum over a large portion of the energy range. Since
other sources also show the limitations of the continuum models near
this energy, and there is little evidence for a third \crsf\ harmonic
in the spectrum, we conclude that this feature is a result of the
continuum model and not a \crsf.

%%
%% 4U 1538-52
%%
\subsection{4U~1538$-$52}\label{subsec:4u1538-52}

The pulsar 4U~1538$-$52 is in an eclipsing system with the B0~Iab
star QV~Nor \citep{rub94,cor93}, with accretion proceeding
through the companion's wind rather than Roche-lobe overflow
\citep{cra78}. The source was first detected with the \uhuru\
satellite \citep{gia74}, with coherent \wsim530\,s pulsations later
found in data from \arielv\ \citep{dav77}. The binary orbital
ephemeris is well known, with an orbital period of 3.7\,d
\citep{cla00}.

A \crsf\ was discovered with the \ginga\ satellite by \citet{cla90}.
The feature, centered at \wsim20\,keV, was found to vary by \wsim15\%
through the pulse.  There is no evidence for a \wsim40\,keV harmonic
in the spectrum, both phase averaged and phase resolved
\citep{mih95,coburn01}. The observation analyzed here (see
Fig.~\ref{fig:specfigstwo}, upper right) spanned nearly the entire
binary orbit of the system (excepting the interval during the \xray\
eclipse). We find a \crsf\ energy of $20.66_{-0.06}^{+0.05}$\,keV in
the source spectrum, with no indication of a second harmonic.

As with 4U~1907+09 (\S\ref{subsec:4u1907+09}), the pattern of fit
residuals below \wsim10\,keV (which arise from the inability of the
model to fit the 10\,keV feature) make the interpretation of the
spectrum in that energy range difficut. While appearing similar to the
residuals seen in 4U~1907+09, the size of this deviation is smaller
(\wpm1.5\%). As with 4U~1907+09, we allowed for another absorption
line at near \wsim10\,keV. However, the fit ``line'' was a broad
feature (\wcy{}=$11_{-2}^{+3}$) with an energy constrained to be less
than 4\,keV, and that acted to modify the entire continuum below
\wsim20\,keV. Therefore we again conclude that the pattern of
residuals is due to the continuum model, and not a \crsf\ in the
spectrum.

%%
%% GX 301-2
%%
\subsection{GX~301$-$2}\label{subsec:gx301-2}

The pulsar GX~301$-$2 (4U~1223$-$62) is a wind fed neutron star in a
41.5 day eccentric ($e=0.47$) orbit around the supergiant companion
Wray 977 \citep{sat86}. Pulsations from GX~301$-$2 were first detected
by \citet{whi76b} at \wsim700\,s. The pulsar undergoes regular flares
\wsim1.4\,d before periastron passage, along with a smaller flare near
apastron \citep{pra01,koh97,pra95}. The primary flare is probably due
to passage through the mass outflow from the companion star which is
directed primarily perpendicular to its rotational axis. The \xray\
spectrum is characterized by a large and variable absorption column
and a strong Fe-K line \citep{lea90,lea89b,whi84}.

There was a suggestion of a \crsf\ in this source at $35.6\pm1.6$\,keV
using observations obtained with the \ginga\ satellite at a reduced
high-voltage, extending the spectral range out to 60\,keV
\citep{mak92,mih95}. But due to uncertainties in the background
subtraction above 36\,keV and limited high energy statistics, the
detection of the \crsf\ was not firm \citep{mak92}. Also, using
\heao-1/A4 data during a source flare, \citet{rot87} reported a
possible absorption line at \wsim24\,keV with an \ftest\ probability
of $10^{-2}$. With the \rxte\ we have been able to confirm the
presence of a \crsf\ in the spectrum, and we find a centroid energy of
$\ecy{}=42.4_{-2.5}^{+3.8}$\,keV. A \crsf\ has also been observed with
the \sax, but at a higher ($49.5\pm1.0$\,keV) energy. \citep{orl01}.

The \rxte\ spectrum in Fig.~\ref{fig:specfigstwo} (lower left) was
obtained during a normal, non-flaring state of GX~301$-$2. The
low-energy absorption measured with the \rxte\ is quite large, with
$\nh=(2.8\pm0.1)\times10^{23}$\,\pcmsq. This is larger than what was
observed with \tenma\ \citep{lea90} and \ginga\ \citep{mih95},
although as mentioned above the low-energy absorption column is known
to be highly variable. It is this large \nh\ that gives rise to the
edge seen in the inferred photon spectrum of the source (see
Fig.~\ref{fig:specfigstwo}, lower left). The \rxte\ spectrum also
exhibits a prominent Fe-K line, with an 1.1\,keV EW. Like other
accreting pulsars there is an indication of a systematic residual
centered near 10\,keV ($\sim\pm0.5$\%).

%%
%% 4U 0352+309
%%
\subsection{4U~0352+309 (X~Per)}\label{subsec:4u0352+309}

The \xray\ source 4U~0352+309 is an unusually low luminosity
($4.2\times10^{34}$\,\lum) pulsar in a system with the Be star
X~Persei (X~Per). It is also unusual in the sense that while most
Be/\xray\ binary systems are transient in nature, 4U~0352+309 is
a source of persistent \xray\ emission. This is perhaps due to
the relatively wide and nearly circular orbit of the system
\citep{del01}. Its \wsim837\,s pulsation period was discovered
with the \uhuru\ satellite \citep{whi76,whi77}, and is still one
of the longest periods of any known accreting pulsar. The
distance to the system, based on optical observations of the
companion X~Per, is 0.95\wpm0.20\,kpc \citep{tel98}.

A \crsf\ at $28.6_{-1.7}^{+1.5}$\,keV was discovered in this source by
\citet{cob01xper}. The spectrum, both here and in the previous work,
is the sum of 40 pointings \citep[see Table 1 of][]{cob01xper}
spanning from 1998 July 1 through 1999 February 27, and the authors
discuss in detail the validity of summing these observations. We found
that, while the spectrum of 4U~0352+309 is not well fit by the
standard pulsar continuum, it can be modeled successfully by a
blackbody plus a power-law modified by a \crsf. In these fits we fixed
the neutral absorption to be $1.5\times10^{21}$\,\pcmsq, based on
measurements from satellites with significant response below 1\,keV
(for example \rosat, \citealt{hab94,mav93}, \bbxrt, \citealt{sch93},
\copernicus, \citealt{mas76}, and \sax, \citealt{sal98}). As in previous
observations of the source we found no evidence of an Fe-K line in the
spectrum, with a 90\% confidence upper limit on the equivalent width
of 13\,eV for a Gaussian line centered at 6.4\,keV and with a 0.5\,keV
sigma.  This is consistent with the \wsim6\,eV upper limit of
\citet{sal98}.

Although the existence of a high energy cutoff in the spectrum is only
suggestive, for this work we used a \plcut\ model in the same way as
we used the \mplcut\ the other sources. Since the cutoff itself was
barely supported by the data, we were unable to apply our smoothing
technique. Since the cutoff itself had a minimal affect on the other
fit parameters, this is unlikely to be a problem for this source.
When we let the \plcut\ model fit, we found cutoff and folding
energies of $\ecut = 57_{-17}^{+12}$ and $\efold = 50_{-30}^{+107}$
respectively. These, along with the other parameters associated with
this fit, are the values listed in Tables~\ref{table:plcutfits} and
\ref{table:params}.

%% /*******************************************************************
%% ** Analysis                                                       **
%% *******************************************************************/

\section{Analysis}\label{sec:fits}

With all of the phase averaged spectra fit in a consistent way with
the same model (except 4U~0352+309 which did not require a high energy
cutoff), we can now discuss any observed correlations and what these
imply for the physical parameters in the pulsar polar regions. The
spectral shape parameterizations are given in
Table~\ref{table:plcutfits}, and the magnetic fields as measured by
their cyclotron lines in Table~\ref{table:bfields}. Of particular
interest is how the \crsf\ parameters (\ecy{}, \wcy{}, \dcy{}) relate
to the continuum (\wgamma, \ecut, \efold), and thus how the standard
pulsar spectrum is affected by the magnetic field.

To identify the major linear correlations in our dataset (the 6 shape
parameters for each of the 10 source spectra), we conducted a
correlation analysis \citep[e.g.][]{bevington92}. The results of which
are given in Table~\ref{table:correlate}. This prescription does not
account for the uncertainties on each point, so these numbers should
merely be used as a guide on where to search in the data. The
coefficient for uncorrelated data is 0, while a perfect
correlation/anticorrelation results in $1$/$-1$ respectively. In
Table~\ref{table:correlate} we see that the primary interdependence is
between the \crsf\ line centroid energy and width (shown in bold) with
a \wsim2\% chance that this is due to a random signal in the data.

There are 2 sources in our analysis whose continua are slightly
different from the other 8 pulsars: 4U~0352+309 and 4U~1626$-$67.
Although 4U~0352+309 has a \crsf\ similar to the rest of the pulsars
presented here, it has a continuum which is quite unlike the standard
pulsar spectral shape. In \citet{cob01xper} we argue that, because of
its very low accretion rate, a thermal component that is normally
buried by accretion is partially revealed. Also, due to inefficient
cooling with lower densities, the break in the spectrum is at a much
higher energy, $\ecut\gtrsim50$\,keV. For 4U~1626$-$67, a pulse phase
resolved spectral analysis reveals that the cutoff energy changes from
\wsim6\,keV to \wsim27\,keV through the pulse \citep[see][]{coburn01}. 
Another reason why these two sources might be different is that in
both cases the binary systems are viewed nearly face on, while 6 of
the remaining 8 are known to be viewed nearly edge on (see
Table~\ref{table:systems}).

While we have not excluded either of these sources from the
overall analysis, when we exclude both 4U~1626$-$67 and
4U~0352+309 from the correlation analysis a new dependency
becomes apparent (see Table~\ref{table:correxcl}). There is
now an association between the \mplcut\ cutoff energy \ecut\ and
the \crsf\ line energy, with a \wsim0.5\% probability of the
correlation being due simply to chance. Additionally, the
connection between the \crsf\ width \wcy{} and energy \ecy{}
becomes tighter, with the same \wsim0.5\% probability of a chance
correlation.

Lastly, in addition to the \wcy{}-\ecy{} and \ecut-\ecy{}
correlations, our analysis also found a new and previously unreported
relation between the relative line width \wcy{}/\ecy{} and the line
depth \dcy{}. This new dependence is discussed in
\S\ref{subsec:depth.vs.width}.

\subsection{Monte Carlo Simulations}\label{sec:monte}

Although care has been taken to ensure that any systematic effects in
the fitting procedure were minimized, other effects might be
responsible for the correlations rather than the actual
spectrum. These problems could be residual systematics inherent in the
\mplcut\ model, selection biases in finding \crsfs\ with satellite
spectrometers, and the phenomenological nature of the continua
models. In particular, both very broad and very narrow
\crsfs\ are very difficult to identify with current instruments.
Since two of our results relied on the \crsf\ width, we performed a
series of Monte Carlo simulations to explore both the fitting process,
and regions of parameter space in which we were sensitive to finding
\crsfs.

\subsubsection{\wcy{} versus \ecy{}}\label{subsec:monte.nrg.vs.width}

Our first simulation was to investigate if the observed \wcy{}-\ecy{}
correlation was a result of systematic correlations between the \crsf\
parameters. To test this we observed how a set of simulated \rxte\
spectra, with known input parameters that formed a grid in the
\wcy{}-\ecy{} plane, were fit using our software and models. If there
were a systematic movement of fit parameters towards a correlation in
the simulations, then the result presented in
\S\ref{subsec:nrg.vs.width} would be suspect. To implement this we
simulated 6 \rxte\ spectra (each with differing depths \dcy{}) at each
point on a 6$\times$6 grid in the \wcy{} (2--10\,kev) versus \ecy{}
(20--45\,keV) plane. The simulated observations were 15\,ks in
duration and of a pulsar with a moderate flux ($F_{\rm
x}=10^{-9}$\,\flux\ in 2--10\,keV). This is a conservative assumption,
allowing fitting systematics and not counting statistics to dominate
the fitting. We fit each of the resulting 216 spectra, and then
plotted \wcy{} versus \ecy{}. As expected there was some movement in
the fit values from the original grid points, however there was no
systematic trend towards a correlation in the data. Therefore it is
safe to conclude that the correlation between \wcy{} and \ecy{} is not
an artifact of how our 10 pulsar spectra were fit or the model used.

The next simulation was done to see how selection biases might account
for the observed correlation. Given the difficulties in identifying
shallow lines, especially at higher energies, a priori one would
expect some bias in the data. Using the same 15\,ks observations, we
simulated 30 spectra at each point on a 11$\times$17 grid in the
\wcy{} (1--12\,keV) versus \ecy{} (16--50\,keV) plane. We fit each of
the resulting spectra initially without a \crsf, and then noted the
change in \wchi\ upon adding one to the model. The dotted line in
Fig.~\ref{fig:nrg.vs.width} shows the contour corresponding to
$\Delta\chi^{2}$ of 9.21 \citep[99\% confidence for two degrees of
freedom,][]{lam76}, with the hatched area indicating regions where a
\crsf\ is indistinguishable from the continuum.

This ``non-detection'' region is easily understood. Broad features,
especially at high energies where the continuum is falling steeply and
statistics are running out, are nearly impossible to identify given
the phenomenological nature of the continuum models. And while the red
wing of broad \crsfs\ at these energies will affect the continuum
where the statistics are better, it again becomes a question of
distinguishing between a real cyclotron feature and the choice of
continuum. Also, narrow lines at all energies are difficult to detect
given the finite energy resolution of not just the \rxte, but
essentially all hard \xray\ satellite instruments so far.

Although the boundary does follow the trend in the data, its exact
position depends on both the source flux and the length of the
observation. As either increases, so will the ability to resolve lines
at higher energies and thus move the dotted line to the
right. Therefore it is unlikely that the absence of narrow lines at
high energies can be completely explained as a selection effect. Also,
selection effects cannot account at all for the lack of broad lines at
lower energies.

\subsubsection{Relative \wcy{} versus \dcy{}}
\label{subsec:monte.depth.vs.width}

To test for systematic effects in the \crsf\ depth versus relative
width correlation, we again performed Monte Carlo simulations. We
first simulated six 15\,ks long \rxte\ spectra with \crsfs\ at various
energies on a $6\times6$ grid in the \wcy{}/\ecy{} (0.05--0.3) versus
\dcy{} (0.1--1.1) plane. We then fit each of the resulting 216 spectra.
Again, although there was scatter in the fit values away from the
initial grid points, there was no systematic movement that could
explain the observed correlation.

To test for a selection bias we next simulated 30 spectra at each
point on a $11\times12$ grid ($0.05\le\wcy{}/\ecy{}\le0.3$,
$0.05\le\dcy{}\le0.8$), each with a \crsf\ at 30\,keV. We fitted each
of the resulting spectra without a \crsf, and noted the improvement
(or lack thereof) in \wchi\ after the addition of an absorption line
to the model. The resulting $\Delta\chi^{2}$ of 9.21
\citep[99\% confidence for two degrees of freedom,][]{lam76} contour
is plotted as a dotted line in Fig.~\ref{fig:depth.vs.fracwidth},
with the hatched regions indicating where the observations were not
sensitive to the presence of a \crsf\ in the spectrum. The region
along the $x$-axis, narrow features at all depths, is simply due to
the finite spectral resolution of the \hexte\ phoswhich detectors. The
region that follows the $y$-axis, shallow features at all widths,
arises from the difficulties in distinguishing shallow \crsfs\ from
the continuum. The dotted line (along with the shaded areas) is meant
as a guide to the eye, and some minor movement is inevitably due to
changes in the \crsf\ energy or shape of the underlying continuum. The
observed correlation between relative width \wcy{}/\ecy{} and depth
\dcy{} cannot, however, be explained merely in terms of either model
biases or selection effects. Therefore it is safe to conclude that
this is a real, physical effect.

%% /*******************************************************************
%% ** Results                                                        **
%% *******************************************************************/

\section{Results}\label{subsec:results}

In this section we discuss the results found in the data analysis of
\S\ref{sec:fits}. These are: correlations between the cyclotron width
\wcy{} and energy \ecy{} (\S\ref{subsec:nrg.vs.width}), between the ratio
\wcy{}/\ecy{} with the cyclotron depth \dcy{}
(\S\ref{subsec:depth.vs.width}), and between the cutoff energy \ecut\
with the cyclotron energy \ecy{} (\S\ref{subsec:nrg.vs.ecut}). At the
end of this section we briefly discuss some possible implications of
the spectral feature near 10\,keV.

\subsection{\wcy{} versus \ecy{}}\label{subsec:nrg.vs.width}

\begin{figure}[H]
\includegraphics[width=3.5in]{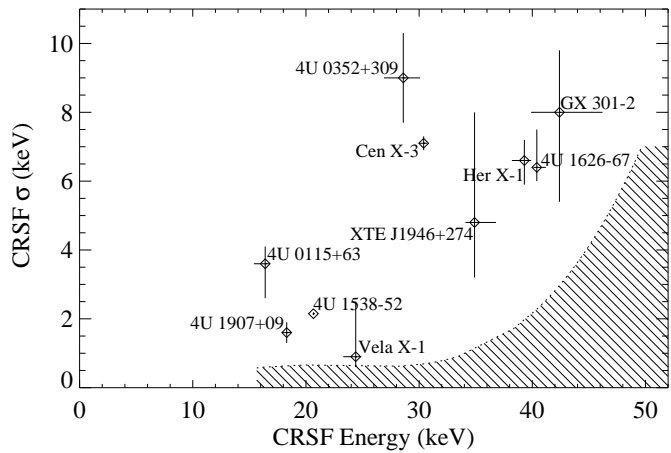}
\caption{\label{fig:nrg.vs.width}
\crsf\ width \wcy{} versus centroid energy \ecy{}. The correlation
between the two parameters is obvious. The shaded region indicates a
non-detection region for \crsfs\ based upon the results of Monte Carlo
simulations (99\% confidence). \crsfs\ in this region would not be
distinguishable from the underlying continua (see,
\S\ref{subsec:monte.nrg.vs.width} ). The doted line is based upon a
conservative estimate of observing time and source flux. Increasing
either of these parameters would allow for the detection of lines at
higher energies, and would therefore move the shaded region to the
right.}
\end{figure}

In Fig.~\ref{fig:nrg.vs.width} we plot the fit values of \crsf\ width
\wcy{} versus energy \ecy{}. This correlation, although not new
\citep{hei99compcrsf,dal00cospar}, is both somewhat expected from
theory and yet still surprising to observe.

The electrons in the scattering region, although quantized
perpendicular are free to move parallel to the magnetic field, and
therefore constitute a 1-dimensional gas \citep{meszaros92}. For cold
electrons the observed \crsf\ width would be nearly a
$\delta$-function. If, however, the electron is moving with a velocity
component $v_{\parallel}$ parallel to the field, then for a photon to
be resonantly scattered it must be at the \crsf\ energy in the
electron rest frame. Therefore, in the laboratory frame the resonance
energy is $E'=\ecy{}[1-(v/c)\cos(\theta)]$, and depends on the photon
propagation angle with respect to the $B$-field. For a
Maxwell-Boltzmann distribution of electrons, the observed \crsf\ FWHM
is given by
\citet{meszaros92}:
\begin{equation}\label{eq:width}
   \eqfwhmpropto
\end{equation}
Where \kte\ is the characteristic electron temperature, \ecy{} the
cyclotron resonance energy, and $\theta$ the viewing angle with
respect to the magnetic field. This gives a very straightforward
dependence of the \crsf\ width \wcy{} on energy
\ecy{} as observed in Fig.~\ref{fig:nrg.vs.width}. There is, however,
a term involving the viewing angle $\theta$. At small angles to the
magnetic field ($\cos(\theta)\sim1$), the motions of the electrons
along the magnetic field lines will result in a maximal Doppler
broadening of the feature, while at large angles ($\cos(\theta)\ll1$)
the effects of thermal broadening are small and the line is expected
to be very narrow.

What is unusual about the correlation seen here is that the systems
should a priori be viewed at \emph{random} values of $\cos(\theta)$,
and thus smear out the correlation (the dependence on \kte, which is
much weaker, is discussed below). Therefore this implies that for
sources where \crsfs\ are observed there is an observational bias
towards a specific viewing angle $\theta$. Interestingly, 6 of the 10
\crsf\ systems discussed here are also among the few accreting
\xray\ pulsars that exhibit eclipses (Her~X-1, Cen~X-3, Vela~X-1, and
4U~1538$-$52) or are thought to be viewed nearly edge on based on fits
to their orbital light curves (GX~301$-$2, 4U~1907+09). Also, since
the angular momentum from the accreted material will seek to align the
neutron star spin with that of the binary plane, knowledge of the
inclination angle of the systems specifies (or nearly so) the pulsar
spin axes. This combination of a selection on a narrow range of
viewing angles $\theta$, with a bias towards systems with a large
inclination, implies that there is a preferred offset angle between
the dipole and spin axes. 

This could be an indication as to why \crsfs\ have not been
observed in sources such as 2S~1417$-$624, GS~1843+00, and
GS~1843$-$024, where observational limitations such as source
brightness and integration length were not a factor
\citep[see][]{coburn01}. The orientation of their binary planes,
and therefore our view of the polar emission regions, are apparently
not favorable to the detection of \crsfs. This also implies that
should the inclination angles of 4U~0115+63, XTE~J1946+274, and
A0535+26 be measured, it will be found that they are observed nearly
edge on.

Given the limited number of sources in our sample and lack of
theoretical work in this area, it is difficult to put an exact number
or even range on what this offset angle between spin and rotation axis
is. In the work of \citet{blu00} for Her X-1, this offset angle was
found to be \wsim18\degree. This would suggest a ``small'' angle, of
less than \wsim30\degree. However, we note that \citet{sco00} find
that for Her X-1 the angle between the spin and $B$-field axes is much
larger (48\degree), although in their model the angle between the
rotation and binary axes is quite large as well (52\degree).

Using a toy model, \citet{wan81} and \citet{wan82} predict that
the angle between the spin and magnetic dipole axis should change
in response to accretion. This is due to a net torque on the
dipole moment resulting from the accreted angular momentum. This
torque will act to align the two axes during episodes of
spin-down, and make them perpendicular during spin-up. The
characteristic time scale for this alignment was estimated to be
$t\sim10^{4}$\,yr, which is similar to the timescale for a
neutron star to come into spin equilibrium with an accretion disk
\citep{els80} and much shorter than the ages of these pulsars.
The authors suggest that most observed accreting \xray\ pulsars
began their lives as rapidly spinning ($\pspin\lesssim1$\,s)
neutron stars, and then experienced significant spin-down due to
a low initial mass transfer and the propeller effect
\citep{ill75}. This is reasonable given that the range of spin
periods for radio pulsars is $10^{-3}-10^{1}$\,s, while for accreting
pulsars it is $10^{-3}-10^{5}$\,s. In this picture there is expected
to be a bias towards small angles between the spin axis and magnetic
moment provided that most neutron stars have not experienced
significant spin-up torques during their lifetimes. This last
assumption is probably not a safe one for most accreting \xray\
pulsars; however, this model does underline the fact that accretion
can affect the orientation of the dipole moment of the neutron star.

If this evolutionary scenario, or indeed any model that seeks to
align the dipole and spin axes, is correct and the angle is small
for most accreting pulsars, then the bias toward edge on systems
can be understood as a bias toward viewing neutron stars
perpendicular to their magnetic dipole moment. If, in general,
the two axes are nearly aligned then the reason for not seeing
\crsfs\ in more sources might simply be due to the distribution
of inclinations of the systems. We note, however, that for any
given pulsar with an unknown inclination angle, there is always
the possibility that the cyclotron energy is out of the effective
\rxte\ detection range.

There are three counter examples to the hypothesis that all pulsars
with observable \crsfs\ are viewed edge on. Both 4U~0352+309 and
4U~1626$-$67 exhibit \crsfs, and yet both binary systems are viewed
nearly face on \citep{del01,cha98}. However, the single peak hard
\xray\ pulsation profiles of 4U~1626$-$67 \citep[e.g.][]{coburn01} and
4U~0352+309 \citep[e.g.][]{del01} are consistent with viewing a single
polar cap that never completely disappears behind the limb of the
neutron star. This might be expected from a source viewed nearly along
the spin axis, and whose magnetic dipole moment is nearly
perpendicular to the spin axis. Under the right circumstances, one
pole could be partially (or even completely) hidden through the pulse
and the other pole viewed nearly edge-on, as with the edge-on sources
in our sample. Still, while this might explain why we observe
\crsfs\ in these sources, it does not address why they might be
orthogonal rotators. It seems unlikely that 4U~0352+309 in particular,
which already has one of the longest spin periods known, was spun-up
from an even longer period as required by the models of \citet{wan81}
and \citet{wan82}.

The third counter example is the eclipsing \xray\ pulsar
OAO~1657$-$415, where so far a \crsf\ has eluded detection. When fit
with the \mplcut\ continuum, the \crsf\ energy implied from the cutoff
energy is \wsim26\,keV (see \S\ref{subsec:nrg.vs.ecut}). A feature at
this energy with a depth greater than $\dcy{}\sim0.2$ can be excluded
at the 3-$\sigma$ level for \crsf\ widths broader than 0.2\ecy{}.

The observed correlation between \wcy{} and \ecy{} is quite broad, and
this might be due in part to the phase averaged spectrum arising from
a range of $\cos(\theta)$ for each source. This effect, while
certainly present, is not expected to be large for two reasons. First,
due to the viewing geometry it is unlikely that the observed range in
$\theta$ for a given source is a full 90\degree. Instead our view
of the polar region will be a smaller subset as the neutron star
rotates. Second, for most of the sources in question the pulsation
profile at the \crsf\ energy is a relatively simple single pulse. As
noted for Cen~X-3, the emission from the peak of this main pulse will
dominate the phase average spectrum, further reducing the effective
range of $\cos(\theta)$ sampled. Future pulse phase resolved fits,
using phases chosen to isolate a specific $\cos(\theta)$, will test
this result and produce a much tighter correlation if
Eq.~\ref{eq:width} is valid.

Assuming that Eq.~\ref{eq:width} is correct, then the existence of the
correlation also indicates a small range of temperatures in the
cyclotron scattering region despite more than 3 orders of magnitude
spread in luminosity (or alternatively, mass accretion rate onto the
neutron star surface). As matter falls onto the polar cap it forms an
accretion structure, where in steady state the amount of matter
falling in balances the matter spreading out at the base. The small
spread in temperatures seen could indicate that the mound parameters
such as area, density profile, etc., naturally find steady state
values that produce a narrow range of electron temperatures in the
scattering region, irrespective of mass accretion rate.

The small range in electron temperatures can also be explained if the
temperature itself is tied to the magnetic field strength. In
numerical simulations by \citet{lam90} of isotropic injection into an
optically thick plasma dominated by cyclotron line cooling and
heating, the temperature is found to vary as $\kte\sim\ecy{}/4$. These
simulations were originally done assuming the hard power-law spectra
of \gammaray\ bursts, and therefore are not directly applicable to the
softer continua of accreting \xray\ pulsars. Still, such a dependence
of electron temperature \kte\ on the magnetic field is consistent with
the observed correlation between the cyclotron energy and width. The
statistics of the data are such that fits to the correlation in
Fig.~\ref{fig:nrg.vs.width} are unable to distinguish between the two
scenarios; with $\wcy{}\propto\ecy{}$ being marginally preferred over
$\wcy{}\propto\ecy{}^{1.5}$ (substituting $\kte\propto\ecy{}$ in
Eq.~\ref{eq:width}), but not at a statistically significant level.

\citet{ara99} cite two caveats to the use of equation
\ref{eq:width} to infer the electron temperature and
viewing angle. The first is that a relativistic treatment of the
scattering produces asymmetric broadening of the line even at
$\theta=90$\degree. Second, Monte Carlo simulations indicate that
the line profiles of the fundamental can be quite complex
\citep{ara00}. The relativistic cross sections for resonant
scattering of photons depend on the angle of photon propagation to the
magnetic field. Since the distribution of scattered photons is
isotropic, the effect of the angular dependence of the cross section
is a non-isotropic angular redistribution of photons.  Although these
simulations preserve the general sense of Eq.~\ref{eq:width}, with
lines at higher energies being broader, these other effects make a
detailed inference of the electron temperature in terms of the \crsf\
width problematic.

\subsection{Relative \wcy{} versus \dcy{}}\label{subsec:depth.vs.width}

\begin{figure}[H]
\includegraphics[width=3.5in]{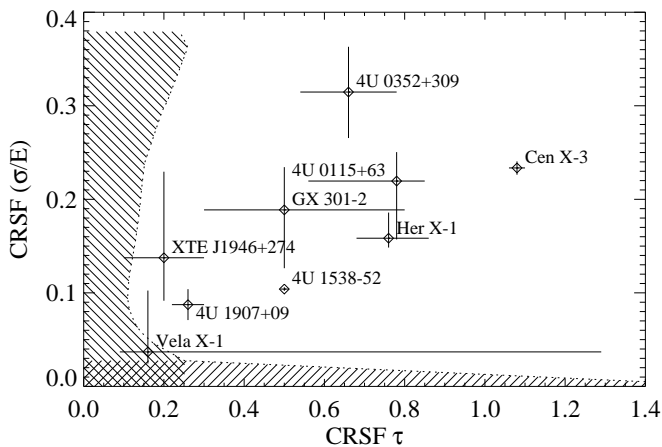}
\caption{\label{fig:depth.vs.fracwidth}
Relative \crsf\ width (\wcy{}/\ecy{}) versus \crsf\ optical depth
\dcy{}. A high degree of correlation is readily apparent in the
data. As in Fig.~\ref{fig:nrg.vs.width}, the shaded regions indicate
regions in parameter space where the \rxte\ is not sensitive to \crsfs\
(99\% confidence, see \S\ref{subsec:monte.depth.vs.width}).}
\end{figure}

In Fig.~\ref{fig:depth.vs.fracwidth} we have plotted the relative
\crsf\ width \wcy{}/\ecy{} versus the optical depth \dcy{}. This is a
much tighter correlation than \wcy{} versus \ecy{}, and has not been
reported before. In \S\ref{subsec:nrg.vs.width} we noted that as the
energies of the lines increased so did their width, which can be
easily understood in terms of Doppler broadening of the lines
themselves. This correlation indicates that as \crsfs\ increase in
depth, the width of the feature as a percentage of the resonance
energy increases as well. Or that as \crsfs\ become deeper, they
become broader.

This observation, that \crsfs\ become relatively broader as they
become deeper is, in fact, \emph{opposite} of what is expected from
the relativistic cross sections alone \citep{ara99}. In the
calculations of the cross sections, resonant photons that travel
perpendicular to the magnetic field have a greater chance of
scattering (and giving rise to a deeper line) than those propagating
along the field. Conversely, off resonance photons traveling parallel
to the $B$-field are more likely to scatter than those
perpendicular. This combination gives cross sections that move from
shallow and broad to narrow and deep as a function of photon
propagation angle. Since the correlation shown here is in the opposite
sense, other factors such as angular redistribution of photons and
photon spawning must be important in the formation of observed
\crsfs. Indeed, in simulations of \citet{ara00}, the predicted shape
of the fundamental \crsf\ can be quite complex and depends greatly on
the magnetic field strength, photon injection and scattering
geometries, and viewing angle. Whether or not the Monte Carlo
simulations can reproduce this result is unclear (R. Araya, private
communication), and needs to be studied further.

It is interesting to note that the shallow fundamental \crsf\ of
Vela~X-1 was until recently considered controversial \citep{kre01}. In
Fig.~\ref{fig:depth.vs.fracwidth} it is on the threshold of
detectability for relatively short integration times (15\,ks in
\hexte\ for the purposes of the simulations), and therefore might
easily be indistinguishable from the continuum in a given
dataset. This could explain why it has been seen in some observations
\citep[e.g.][]{mih95,kre97,kre99,kre01} and not in others
\citep[e.g.][]{orl98vela}.

\subsection{\ecut\ versus \ecy{}}\label{subsec:nrg.vs.ecut}

\begin{figure}[H]
\includegraphics[width=3.5in]{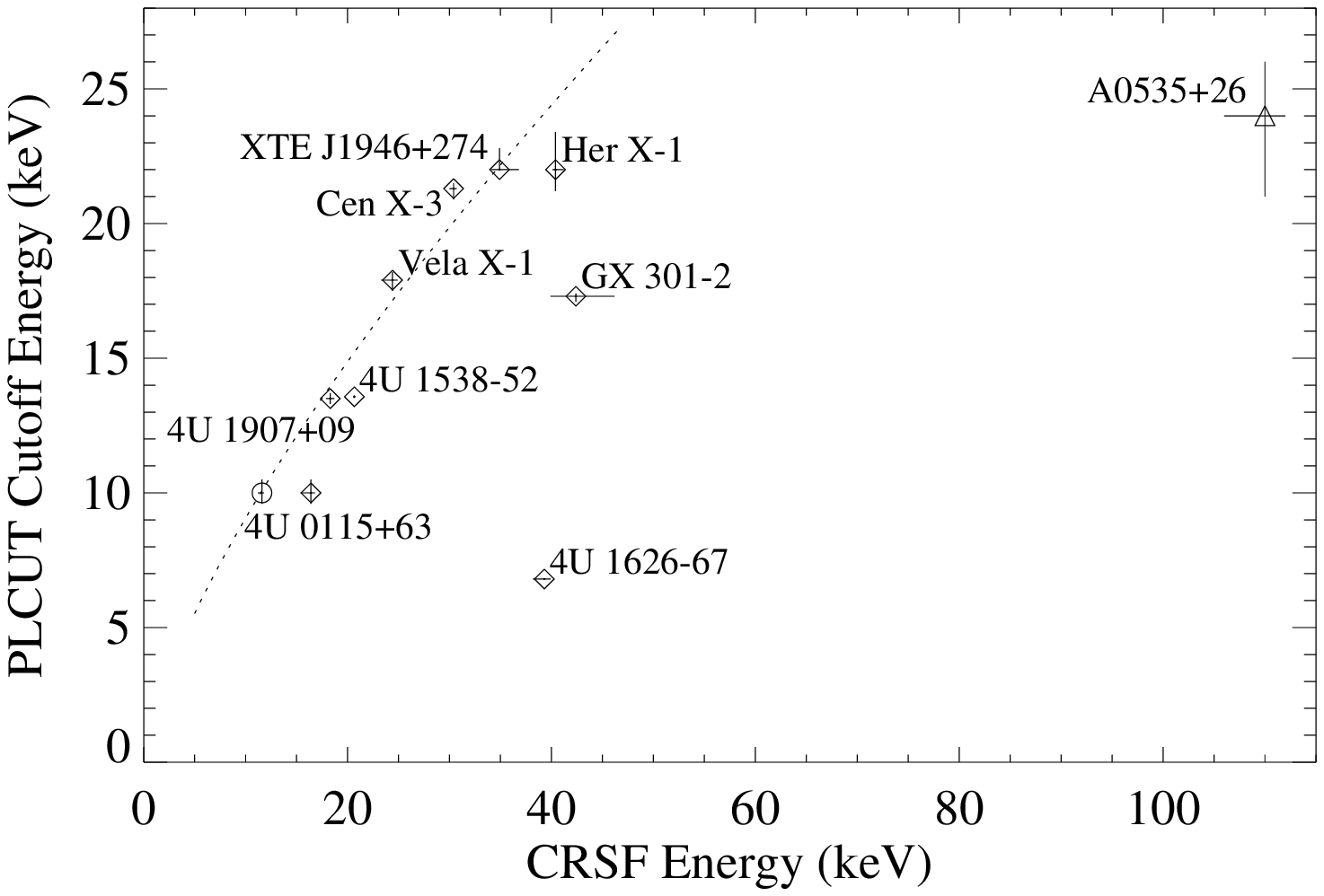}
\caption{\label{fig:nrg.vs.cut}
\mplcut\ cutoff energy \ecut\ versus \crsf\ centroid energy \ecy{}. Two
\crsf\ energies are shown for 4U~0115+63, the diamond indicating the
fit value (and therefore the center of the feature as modeled with a
\gabs) and a circle at $1/2$ the energy of the second harmonic (and
thus a is a measure of $B$-field, see \S\ref{subsec:4u0115+63} for
details). The \cgro/\osse\ energy \citep{gro95} and \hexe\ cutoff
\citep{ken94} of A0535+26 are shown as a diamond, while
the lower limit for a cutoff energy in 4U~0352+309 is outside the
plot. The new \crsf\ source XTE~J1946+47, along with Cen~X-3, Her~X-1,
and A0535+26 indicate a flattening of the power-law correlation (shown
as a dotted line) observed by \citet{mak99}. }
\end{figure}

Our third result is a relationship between the cutoff energy \ecut\
and the \crsf\ energy \ecy{} (see Fig.~\ref{fig:nrg.vs.cut}). A
relationship between the two parameters was first noticed by
\citet{mak92} and later refined by \citet{mak99}. There, however, the
authors used \crsf\ energies obtained with several different
instruments and derived using a variety of continuum models. To find
the \plcut\ cutoff energies, the authors used fits to \ginga\
observations without a \crsf, and compared them to values of the
\crsf\ energy found in the literature often from non-contemporaneous
observations. They found that the relationship was consistent with a
power law, $\ecut\propto\ecy{}^{0.7}$, indicating a saturation as
compared to a linear correlation. Our work has the advantage that each
point is derived from a uniform model fit to a single spectrum. Also,
with the exception of A0535+26, all points are from the \rxte\
instruments.

As indicated by the dotted line in Fig.~\ref{fig:nrg.vs.cut} which
represents $\ecut{}\propto\ecy{}^{0.7}$, our results confirm the
relationship for \crsf\ energies below \wsim35\,keV. Above 35\,keV,
however, there is an indication of a change in slope in the
correlation. Specifically, the cutoff energies of Cen~X-3,
XTE~J1946+274, Her~X-1, and A0535+26 are the same within errors. If
real, this flattening of \ecut\ is more abrupt than the rather smooth
turnover of the $\ecy{}^{0.7}$ power law. It also suggests that the
relation between the cyclotron and cutoff energies can instead be
described by two linear relations, one below \wsim35\,keV and one
above.

While the complex correlation between the \crsf\ energy and the cutoff
energy implies that the spectral break is related to the $B$-field, it
is not necessarily a magnetic effect. A similar yet physically
distinct possibility is that the break is a function of another
physical parameter, such as the electron temperature \kte. It may be
this intermediate quantity that is proportional to the magnetic field,
up to a saturation point.

Another interesting observation is that the change in slope occurs
at \wsim35\,keV, where the cyclotron energy is \wsim7\% of the rest
mass of the electron. The roll over might be an indication that
relativistic effects in the creation of the continuum, are beginning
to become important at those fields. This is, however, highly
speculative. Whatever the cause, future models of accreting
\xray\ pulsar spectra will need to reproduce this correlation between
the cyclotron line depth and relative width.

For illustrative purposes we have included the results of fits to
\cgro/\osse\ data for the source A0535+26 \citep[indicated with a
triangle in Fig.~\ref{fig:nrg.vs.cut},][]{gro95}. Due to the high
energy threshold of \osse, for this this fit the continuum was modeled
with a \plcut\ with \ecut\ set to zero. Therefore, we have used the
cutoff energy observed by \citet{ken94} with \hexe. Both \citet{gro95}
and \citet{ken94} find a suggestion of a feature near \wsim55\,keV,
although as was the case for Vela~X-1 the existence of this as the
fundamental is considered somewhat controversial. If the fundamental
is at 55\,keV then this is consistent with the observed correlation,
while if it is at 110\,keV then it significantly strengthens the
already apparent saturation.

Not shown in this plot is the source 4U~0352+309, which has a cutoff
energy lower limit of $\ecut=60$\,keV. However, as we mentioned in
\S\ref{subsec:4u0352+309}, the spectrum of 4U~0352+309 is
unusual when compared to other accreting \xray\ pulsars. The
pulse phase resolved spectrum of 4U~1626$-$67 is different as
well. In particular, the cutoff energy \ecut\ is a strong
function of pulse phase, changing by a factor of 4.5 through the
pulse. This change in the cutoff energy as a function of rotation
phase is unlike what is seen in other sources, and therefore it
is not surprising that the phase average value for this source
does not fall on the observed correlation.

The third point that lies off of the correlation is GX~301$-$2. It is
interesting to notice, however, that there is less discrepancy if the
fundamental \crsf\ lies \emph{not} at 42\,keV, but instead at 21\,keV
as suggested by \citet{rot87}. This may be an indication that the
fundamental feature, as in Vela~X-1, is shallow and difficult to
detect.  Since the \crsf\ parameters of energy, width, and depth are
known to change with pulse phase, a full pulse phase resolved analysis
could answer this question. Unfortunately the archival \hexte\ data
was obtained in a mode where such an analysis is impossible. Recent
\rxte\ observations, however, should answer this question
(I. Kreykenbohm, private communication).

Since we are inferring the effect of the magnetic field on the
standard pulsar continuum, we have plotted two values of 4U~0115+63
\crsf\ centroid energy. The higher value (plotted as a diamond) is the
result of the parameterization used here, and is the center of the
absorption feature as modeled with the \gabs\ analytic form. The
second (plotted as a circle) is at half the energy of the second
harmonic, and therefore is a better estimate of the pulsar magnetic
field (see \S\ref{subsec:4u0115+63}). We did not make this distinction
in the previous 2 sections because those were investigations into the
\crsf\ shapes themselves, and the shape of the \crsf\ in 4U~0115+63 as
parameterized with the \gabs\ model is centered at \wsim16\,keV. Here,
however, since the \crsf\ energy is used to measure the $B$-field the
distinction is necessary.

\subsection{The Feature at 10\,keV}\label{subsec:10keV}

One interesting feature in most of these \mplcut\ fits is the wiggle,
or ``bump,'' in the spectrum in the range 8--12\,keV. It might easily
be mistaken for a \crsf, especially given the size of the residuals
in, for example, the spectrum of 4U~1907+09 (see
Fig.~\ref{fig:specfigstwo}, upper left). However, these features
cannot be fit with simple absorption line models. This wiggle is
evident in the spectra of many accreting \xray\ pulsars, including
those that do not otherwise exhibit a \crsf\ \citep[e.g. GS~1843+00,
see][]{coburn01}. While it often appears as a simple dip, it can also
evidence itself as a pure bump (e.g. Her~X-1, see
\S\ref{subsec:herx1}) or a pseudo-P-Cygni profile. Since this bump is
consistently at or near 10\,keV, irrespective the \crsf\ energy, the
feature is probably not a magnetic effect.

The bump is not just apparent in \rxte\ spectra either. In
\citet{mih95}, the residuals in fits using the \npex\ model to
\ginga\ observations (e.g. 4U~1538$-$52, 4U~1907+09, and V0331+53)
show a systematic structure around 10\,keV \citep{mih95}. Pulse phase
resolved spectroscopy indicates that the feature changes through the
pulse. This \wsim10\,keV bump is also evident in the residuals seen
with the \sax\ High Pressure Gas Scintillation Proportional Counter
(HPGSPC) in Cen~X-3 \citep{san98}. Since this feature has been seen in
a number of sources by several satellites, and yet is not present in
the observed spectra of other sources (like the Crab nebula/pulsar) it
is safe to conclude that it is an inherent feature in the spectra of
accreting pulsars.

The shape of the feature seen in plots of residuals almost certainly
arises from modeling the continuum with a simple power-law (\mplcut) or
nearly a power-law (\fdco) below the cutoff energy. This improper
modeling may also account for the different patterns of residuals that
are observed, rather than physical changes in the conditions creating
the feature. The \npex\ model, with two components that each have an
independent normalization, can mimic this feature somewhat. Still, as
seen in \citet{mih95}, the model is only marginally successful at
accounting for the feature for most sources.

While it could be safely ignored in previous missions, the large
collecting area of the \pca, which is 50\% larger than the previously
largest proportional counter flown (the \ginga\ LAC), has made this
spectral ``feature'' something that can no longer be ignored. As the
collecting areas of satellite spectrometers become larger, this
problem will continue to hinder accurate spectral modeling. Therefore
we strongly encourage the development of theoretical models for
accreting \xray\ pulsar spectra that can be compared directly with the
observations and reduce the reliance on the currently used
phenomenological models.

%% /*******************************************************************
%% ** Summary                                                        **
%% *******************************************************************/

\section{Summary}\label{sec:summary}

There are three principle results of this class analysis of the
spectra and cyclotron features of accreting \xray\ pulsars. The first
two results involve correlations among the shape parameters of the
\crsfs\ themselves. We find that the observed \crsf\ widths are
roughly proportional to their energy. If the widths are primarily due
to Doppler broadening, then this implies a viewing angle selection
bias in finding sources that exhibit cyclotron features. Since 6 of
the 10 sources discussed here are in systems that are viewed nearly
edge on, a selection bias on viewing angles further suggests a
preferred offset angle between the dipole and spin axes of the neutron
star.

The next result is that, at least for the fundamental features, deeper
\crsfs\ are also broader, even when scaled by the centroid
energy. This is difficult to understand simply in terms of the
relativistic cross sections alone, which are in the
\emph{opposite} sense. This implies that other effects, such as
photon spawning or the non-isotropic angular redistribution of
photons, are important and should continue to be considered in
theoretical efforts.

Lastly, we find a correlation between the magnetic field strength
and the spectral cutoff energy. The existence of a correlation
indicates that the observed spectral break is either a magnetic
effect, or perhaps is tied to the magnetic field through some
intermediate quantity. In the correlation itself there is a
departure from the power-law observed by \citet{mak99}, either a
roll over or break in the slope, near a cyclotron resonance
energy of 35\,keV. This might be an indication that the break is
indeed tied to another quantity, such as the electron
temperature, that is then tied to the magnetic field up to a
saturation point.

We also discussed a departure in observed pulsar spectra from the
standard pulsar continuum shape. While the standard shape is
still applicable over most of the hard \xray\ band, modern
satellites have shown that it is inadequate near 10\,keV. This
highlights the need for theoretical work in how the \xray\
continuum in these pulsars is formed, and when fitting spectra a
departure away from the purely phenomenological models currently
being used.

%% /*******************************************************************
%% ** Acknowledgements                                               **
%% *******************************************************************/

\acknowledgments This work was supported by NASA grant NAS5-30720,
NSF Travel Grant NSF INT-9815741, LTSA Grant NAG5-10691, and DAAD
travel grants.

%% /*******************************************************************
%% ** Bibliography                                                   **
%% *******************************************************************/

%% /*******************************************************************
%% ** Figures                                                        **
%% *******************************************************************/

%% /*******************************************************************
%% ** Tables                                                         **
%% *******************************************************************/
\begin{table}[H]
\begin{minipage}{\textwidth}
\caption[\rxte\ \crsf\ Sources]%%
{\label{table:systems} \rxte\ \crsf\ Sources}
\begin{tabular}{lccccc} \hline \hline
 System         & Type  & \pspin & \porb  & Companion          & Eclipsing  \\
                &       & (s)    & (days) & (MK Type)          & \\ \hline
 Hercules~X-1   & \lmxb\footnote{Low Mass \xray\ Binary}
                        & 1.2377 & 1.70   & HZ Her (A/F)       & \textbf{Yes} \\
 4U~0115+63     & \bet  & 3.61   & 24.3   & V635 Cas (Be)      & No \\
 Centaurus~X-3  & \hmxb\footnote{High Mass \xray\ Binary}
                        & 4.82   & 2.09   & V779 Cen (O6-8f)   & \textbf{Yes}\\
 4U~1626$-$67   & \lmxb & 7.67   & 0.0289 & KZ TrA             & No \\
 XTE~J1946+274  & \tuc  & 15.83  & --     & --                 & No \\
 Vela~X-1       & \hmxb & 283.2  & 8.96   & HD 77581 (B0.5 Ib) & \textbf{Yes} \\
 4U~1907+09     & \hmxb & 440.4  & 8.38   & (Be?)              & \textbf{Nearly} \\
 4U~1538$-$52   & \hmxb & 528.8  & 3.73   & QV Nor (B0 Iab)    & \textbf{Yes}\\
 GX~301$-$2     & \hmxb & 681    & 41.5   & Wray 977 (B1.5 Ia) & \textbf{Nearly}\\
 4U~0352+309    & \bep  & 837.7  & 250.3  & X Per (09 III-Ve)  & No\\ \hline
\end{tabular}
\end{minipage}
\end{table}

\begin{table}[H]
\caption[\pca\ Configurations During Observations]%%
{\label{table:pcaobs} \pca\ Configurations During
Observations (see \S\ref{sec:rxte}) }
\begin{tabular}{lcccc} \hline \hline
Source        & Gain Epoch & Background Model & \pcus\    & Layers \\
\hline
Her~X-1       & 3     & \skyvle          & 0,1,2     & Top \\
4U~0115+63    & 3     & \skyvle          & 0,1,2,3   & Top \\
Cen~X-3       & 3     & \skyvle          & All       & Top \\
4U~1626$-$67  & 1     & \faint           & All       & Top \\
XTE~J1946+274 & 3     & \skyvle          & 0,1,2     & Top \\
Vela~X-1      & 3     & \skyvle          & 0,1,2,4   & Top \\
4U 1907+09    & 1     & \faint           & All       & All \\
4U 1538$-$52  & 3     & \skyvle/\faint   & 0,1,2     & Top \\
GX~301$-$2    & 3     & \skyvle          & All       & Top \\
4U~0352+309   & 3     & \skyvle          & 0,1,2     & Top \\
\hline
\end{tabular}
\end{table}

\begin{table}[H]\small
\caption[\crsf\ Fits to Different continua]
{\label{table:crsfcont} \crsf\ Fits to Different continua}
\begin{tabular}{l|l|ccc} \hline \hline
 Source        & Continua & \ecy{} & \wcy{} & \dcy{} \\ \hline
 Her~X-1       & \mplcut  & $40.4_{-0.3}^{+0.8}$ & $6.4_{-0.4}^{+1.1}$  & $0.76_{-0.08}^{+0.10}$ \\
               & \fdco    & $41.4_{-0.3}^{+1.5}$ & $9.2_{-0.1}^{+0.6}$  & $1.69_{-0.07}^{+0.66}$ \\
 Cen~X-3       & \mplcut  & $30.4_{-0.4}^{+0.3}$ & $7.1\pm0.2$          & $1.1\pm0.1$ \\
               & \fdco    & $29.9\pm0.1$         & $8.8\pm0.1$          & $1.4\pm0.1$ \\
 XTE~J1946+274 & \mplcut  & $34.9_{-0.8}^{+1.9}$ & $4.8_{-1.6}^{+3.2}$  & $0.2\pm0.1$ \\
               & \fdco    & $35.7_{-1.1}^{+1.2}$ & $3.7\pm1.1$          & $0.3\pm0.1$ \\
 GX~301$-$2    & \mplcut  & $42.4_{-2.5}^{+3.8}$ & $8.0_{-2.6}^{+1.8}$  & $0.5_{-0.2}^{+0.3}$ \\
               & \fdco    & $39.8_{-3.4}^{+0.2}$ & $10.4_{-4.6}^{+0.1}$ & $0.4\pm0.1$ \\
\hline
\end{tabular}
\end{table}

\begin{table}[ht]
\caption[Summary of Observations]%%
{\label{table:obs} Summary of Observations}
\begin{minipage}{\textwidth}
\begin{tabular}{lclccccc} \hline \hline
    && & \multicolumn{2}{c}{PCA}
     & & \multicolumn{2}{c}{HEXTE}
       \\ \cline{4-5} \cline{7-8}
Source && Date & Livetime & Rate\footnote{In the range 3--20\,keV}
       && Livetime & Rate\footnote{In the range 15--100\,keV} \\
       &&      & (ks)     & (\cpspcu)
       && (ks) & (\cpspcu) \\
\hline
Her~X-1       && 1997 Sep 14          & 19.4  & $254.2\pm0.2$  &&  11.9 & $56.2\pm0.1$\\
4U~0115+63    && 1999 Mar 3           & 16.7  & $927\pm1$      &&  15.4 & $112.6\pm0.1$\\
Cen~X-3       && 1997 Feb 28/Mar 2    &157.3  & $560.4\pm0.4$  && 129.4 & $51.83\pm0.04$\\
4U~1626$-$67  && 1996 Feb 11--14      &121.5  & $26.78\pm0.02$ &&  81.8 & $6.49\pm0.05$\\
XTE~J1946+274 && see \citet{hei01xtej}& 30.6  & $237.1\pm0.1$  &&  19.7 & $43.0\pm0.1$ \\
Vela~X-1      && 1998 Jan 22          & 19.1  & $208.0\pm0.3$  &&  11.5 & $120.4\pm0.2$\\
4U~1907+09    && 1996 Feb 21          & 18.3  & $18.29\pm0.04$ &&  12.2 & $1.9\pm0.2$  \\
4U~1538$-$52  && 1997 Jan 3--5        & 100.8 & $46.83\pm0.04$ &&  62.1 & $5.10\pm0.06$\\
GX~301$-$2    && 1996 Sep 12          & 27.0  & $114.6\pm0.08$ &&  20.6 & $50.1\pm0.1$\\
4U~0352+309   && see \citet{cob01xper}& 161.4 & $20.55\pm0.03$ && 109.3 & $1.60\pm0.04$ \\
\hline
\end{tabular}
\end{minipage}
\end{table}

\begin{table}[ht]
\caption[Source \mplcut\ Parameterizations]%%
{\label{table:plcutfits} Source \mplcut\ Parameterizations}
\begin{minipage}{\textwidth}
\begin{tabular}{lccccccc} \hline \hline
 Source & \wgamma
        & \ecut 
        & \efold
        & \ecy{}
        & \wcy{}
        & \dcy{}
        & \wchired/DOF \\ \hline
  Her X-1  & $0.93_{-0.02}^{+0.01}$
           & $22.0_{-0.8}^{+1.4}$
           & $10.8_{-0.3}^{+0.2}$
           & $40.4_{-0.3}^{+0.8}$   
           & $6.4_{-0.4}^{+1.1}$ 
           & $0.76_{-0.08}^{+0.10}$
           &  0.691/76 \\
 4U~0115+63 & $0.26_{-0.02}^{+0.10}$ 
            & $10.0_{-0.4}^{+0.5}$ 
            & $9.3_{-0.2}^{+0.5}$
            & $16.4_{-1.0}^{+0.4}$   
            & $3.6_{-1.0}^{+0.5}$  
            & $0.78_{-0.22}^{+0.07}$
            & 1.105/83 \\
 Cen~X-3   & $1.24_{-0.02}^{+0.01}$ 
           & $21.3_{-0.4}^{+0.2}$ 
           & $6.67_{-0.06}^{+0.09}$
           & $30.4_{-0.4}^{+0.3}$  
           & $7.1\pm0.2$         
           & $1.08\pm0.02$
           & 1.536/114\\
 4U~1626$-$67 & $0.88\pm0.05$
              & $6.8\pm0.3$          
              & $38.8_{-4.6}^{+5.5}$
              & $39.3_{-1.1}^{+0.6}$  
              & $6.6_{-0.7}^{+0.6}$  
              & $2.1\pm0.2$
              & 1.263/122  \\
 XTE~J1946+274 & $1.14_{-0.03}^{+0.02}$
               & $22.0_{-0.9}^{+0.8}$ 
               & $8.3_{-0.4}^{+0.5}$
               & $34.9_{-0.8}^{+1.9}$   
               & $4.8_{-1.6}^{+3.2}$  
               & $0.2\pm0.1$
               & 1.067/52   \\
 Vela~X-1   & $0.00_{-0.01}^{+0.09}$
            & $17.9_{-0.4}^{+0.3}$ 
            & $8.8_{-0.1}^{+0.2}$
            & $24.4_{-1.1}^{+0.5}$   
            & $0.9_{-0.8}^{+0.9}$  
            & $0.16_{-0.07}^{+1.13}$
            & 0.879/64 \\
 4U 1907+09 & $1.236_{-0.012}^{+0.007}$ 
            & $13.5\pm0.2$ & $9.8\pm0.6$
            & $18.3\pm0.4$              
            & $1.6\pm0.2$   
            & $0.26\pm0.04$
            & 1.827/95 \\
 4U 1538$-$52 & $1.161_{-0.014}^{+0.003}$
              & $13.57_{-0.05}^{+0.04}$ 
              & $11.9\pm0.1$
              & $20.66_{-0.06}^{+0.05}$   
              & $2.15\pm0.04$           
              & $0.50\pm0.01$
              & 2.207/116 \\
 GX~301$-$2 & $-0.02\pm0.04$
            & $17.3_{-0.2}^{+0.1}$ 
            & $6.7_{-0.3}^{+0.7}$
            & $42.4_{-2.5}^{+3.8}$
            & $8.0_{-2.6}^{+1.8}$  
            & $0.5_{-0.2}^{+0.3}$
            & 1.240/60 \\
 4U~0352+309 & $1.82\pm0.02$
             & $57_{-17}^{+12}$\footnote{Not required by fit}
             & $50_{-30}^{+107}$$^{a}$
             & $28.6_{-1.7}^{+1.5}$ 
             & $9.0\pm1.3$     
             & $0.66\pm0.12$
             & 0.876/80 \\ \hline
\end{tabular}
\end{minipage}
\end{table}

\begin{table}[H]
\caption[\mplcut\ Other Fit Parameters]
{\label{table:params} \mplcut\ Other Fit Parameters}
\begin{minipage}{\textwidth}
\begin{tabular}{lccccccc} \hline \hline
               &
               & 
               &
               &
               & \multicolumn{3}{c}{``Smoothing'' Fumction} \\ \cline{6-8}
 Source        & \nh ($10^{22}$ \pcmsq)
               & $kT$ (keV) 
               & $E_{FeK}$ (keV)
               & $EW_{FeK}$ (eV)
               & $E$ (keV)  
               & $\sigma$ (keV)
               & $\tau$ \\ \hline
 Her X-1       & -
               & -
               & $6.0_{-0.4}^{+0.2}$/$6.49_{-0.06}^{+0.03}$ 
               & 370/78
               & $22.0_{-0.9}^{+1.3}$
               & $3.0_{-0.2}^{+0.3}$  
               & $0.18_{-0.03}^{+0.05}$ \\
 4U 0115+63    & -                      
               & -
               & -
               & -
               & $11.2_{-0.3}^{+0.4}$
               & $2.2\pm0.3$      
               & $0.37_{-0.12}^{+0.08}$ \\
 Cen X-3       & $2.1\pm0.2$
               & -
               & $6.51\pm0.04$                     
               & 141
               & $21.0_{-0.4}^{+0.3}$  
               & $0.4\pm0.4$           
               & $3.8_{-0.4}^{+0.1}$ \\
 4U 1626-67\footnote{Smoothing function did not significantly affect fit}
               & $1.6_{-0.7}^{+0.9}$ 
               & $0.35_{-0.05}^{+0.04}$
               & -
               & -
               & $6.8\pm0.3$
               & $0_{-0}^{+\infty}$
               & $0_{-0}^{+\infty}$ \\
 XTE J1946+274 & -
               & -
               & -                    
               & -
               & $0.30_{-0.06}^{+0.07}$
               & $22.0_{-1.1}^{+1.1}$   
               & $3.4_{-0.6}^{+0.3}$ \\
 Vela X-1      & -
               & -
               & -                    
               & -
               & $0.28_{-0.03}^{+0.01}$
               & $18.5_{-0.4}^{+0.2}$
               & $2.4\pm0.02$ \\
 4U 1907+09$^{a}$
               & $2.4\pm0.1$
               & -
               & $6.52\pm0.07$
               & 175
               & $13.6_{-0.3}^{0.7}$
               & $0_{-0}^{+0.3}$  
               & $0_{-0}^{+\infty}$ \\
 4U 1538-52$^{a}$
               & $1.98_{-0.18}^{+0.04}$
               & -
               & $6.25\pm0.06$
               & 61
               & $13.6_{-0.2}^{0.3}$
               & $0_{-0}^{+\infty}$
               & $0_{-0}^{+\infty}$ \\
 GX 301-2      & $28\pm1$
               & -
               & $6.445_{-0.005}^{+0.007}$
               & 1100
               & $17.6\pm0.3$
               & $2.30_{-0.12}^{+0.07}$ 
               & $0.27\pm0.02$ \\
 4U 0352+309\footnote{High energy cutoff not required by fit}
               & 0.15\footnote{Fixed}
               & $1.44\pm0.01$
               & -
               & -
               & -
               & -
               & - \\
\hline
\end{tabular}
\end{minipage}
\end{table}

\begin{table}[H]
\caption[Pulsar Magnetic Fields]%%
{\label{table:bfields} Pulsar Magnetic Fields}%%
\begin{minipage}{\textwidth}
\begin{tabular}{lccccc} \hline \hline
 System        & \ecy{}\,keV
               && $B(1+z)^{-1}\times10^{12}$\,G
               && $B\times10^{12}$\,G
 \footnote{Assuming $\rns=10$\,km, $\mns=1.4$\,\msol, and scattering near the
 surface of the star} \\ \hline
 Hercules~X-1   & $40.4_{-0.3}^{+0.8}$    && 3.5 && 4.5 \\
 4U~0115+63\footnote{Using 1/2 the energy of the second harmonic, see text}
                & $11.6_{-0.4}^{+0.2}$    && 1.0 && 1.3 \\
 Centaurus~X-3  & $30.4_{-0.4}^{+0.3}$    && 2.6 && 3.4 \\
 4U~1626$-$67   & $39.3_{-1.1}^{+0.6}$    && 3.4 && 4.4 \\
 XTE~J1946+274  & $34.9_{-0.8}^{+1.9}$    && 3.0 && 3.9 \\
 Vela~X-1       & $24.4_{-1.1}^{+0.5}$    && 2.1 && 2.7 \\
 4U~1907+09     & $18.3\pm0.4$            && 1.6 && 2.1 \\
 4U~1538$-$52   & $20.66_{-0.06}^{+0.05}$ && 1.8 && 2.3 \\
 GX~301$-$2     & $42.4_{-2.5}^{+3.8}$    && 3.7 && 4.8 \\
 4U~0352+309    & $28.6_{-1.7}^{+1.5}$    && 2.5 && 3.2 \\ \hline
\end{tabular}
\end{minipage}
\end{table}

\begin{table}[H]
\caption[Correlation Matrix]%%
{\label{table:correlate} Correlation Matrix, All Sources}
\begin{tabular}{l|cccccc} \hline \hline
        & \wgamma & \ecut & \efold & \ecy{} & \wcy{} & \dcy{} \\ \hline
 \wgamma&  $1.00$ &  $0.56$ & $0.51$ & $-0.08$ & $0.28$       &  $0.12$ \\
 \ecut  &         &  $1.00$ & $0.57$ &  $0.09$ & $0.53$       & $-0.21$ \\
 \efold &         &         & $1.00$ &  $0.15$ & $0.49$       &  $0.50$ \\
 \ecy{} &         &         &        &  $1.00$ & $\mbf{0.70}$ &  $0.37$ \\
 \wcy{} &         &         &        &         & $1.00$       &  $0.46$ \\
 \dcy{} &         &         &        &         &              &  $1.00$ \\ \hline
\end{tabular}
\end{table}

\begin{table}[H]
\caption[Correlation Matrix, Excluding 4U~1626$-$67 and
4U~0352+309]%%
{\label{table:correxcl} Correlation Matrix, Excluding
4U~1626$-$67 and 4U~0352+309}%%
\begin{tabular}{l|cccccc} \hline \hline
        & \wgamma & \ecut & \efold & \ecy{} & \wcy{} & \dcy{} \\ \hline
 \wgamma&  $1.00$ &  $0.25$ &  $0.31$ & $-0.11$      & $-0.02$      &  $0.17$ \\
 \ecut  &         &  $1.00$ & $-0.35$ & $\mbf{0.78}$ &  $0.53$      &  $0.08$ \\
 \efold &         &         &  $1.00$ & $-0.41$      & $-0.57$      & $-0.17$ \\
 \ecy{} &         &         &         &  $1.00$      & $\mbf{0.81}$ &  $0.12$ \\
 \wcy{} &         &         &         &              &  $1.00$      &  $0.60$ \\
 \dcy{} &         &         &         &              &              &  $1.00$ \\ \hline
\end{tabular}
\end{table}


\begin{thebibliography}{}

\bibitem[\protect\astroncite{Araya \& Harding}{1999}]{ara99}
Araya, R.~A., \& Harding, A.~K.,  1999, ApJ, 517, 334

\bibitem[\protect\astroncite{Araya-G{\'o}chez \& Harding}{2000}]{ara00}
Araya-G{\'o}chez, R.~A., \& Harding, A.~K.,  2000, ApJ, 544, 1067

\bibitem[\protect\astroncite{{Arnaud}}{1996}]{arn96}
{Arnaud}, K.~A.,  1996,
\newblock in ASP Conf. Ser. 101: Astronomical Data Analysis Software and
  Systems V, ed. G.~H. Jacoby, J. Barnes, Vol.~5, ~17

\bibitem[\protect\astroncite{Bevington \& Robinson}{1992}]{bevington92}
Bevington, P.~R., \& Robinson, D.~K.,  1992,
\newblock Data reduction and error analysis for the physical sciences,
\newblock  (New York: McGraw-Hill), 2nd edition

\bibitem[\protect\astroncite{Bildsten et~al.}{1997}]{bil97}
Bildsten, L., et~al., 1997, ApJS, 113, 367

\bibitem[\protect\astroncite{{Blum} \& {Kraus}}{2000}]{blu00}
{Blum}, S., \& {Kraus}, U.,  2000, ApJ, 529, 968

\bibitem[\protect\astroncite{{Burderi} et~al.}{2000}]{bur00}
{Burderi}, L., {Di Salvo}, T., {Robba}, N.~R., {La Barbera}, A., \&
  {Guainazzi}, M.,  2000, ApJ, 530, 429

\bibitem[\protect\astroncite{Campana, Israel \& Stella}{1999}]{cam99}
Campana, S., Israel, G., \& Stella, L.,  1999, A\&A, 352, L91

\bibitem[\protect\astroncite{Chakrabarty}{1998}]{cha98}
Chakrabarty, D.,  1998, ApJ, 492, 342

\bibitem[\protect\astroncite{{Chitnis} et~al.}{1993}]{chi93}
{Chitnis}, V.~R., {Rao}, A.~R., {Agrawal}, P.~C., \& {Manchanda}, R.~K.,  1993,
  A\&A, 268, 609

\bibitem[\protect\astroncite{{Clark}}{2000}]{cla00}
{Clark}, G.~W.,  2000, ApJ, 542, L131

\bibitem[\protect\astroncite{{Clark} et~al.}{1990}]{cla90}
{Clark}, G.~W., {Woo}, J.~W., {Nagase}, F., {Makishima}, K., \& {Sakao}, T.,
  1990, ApJ, 353, 274

\bibitem[\protect\astroncite{Coburn}{2001}]{coburn01}
Coburn, W.,  2001,
\newblock {\em Ph.D. thesis\/}, University of California, San Diego, La Jolla,
  CA

\bibitem[\protect\astroncite{Coburn et~al.}{2001}]{cob01xper}
Coburn, W., Heindl, W.~A., Gruber, D.~E., Rothschild, R.~E., Staubert, R.,
  Wilms, J., \& Kreykenbohm, I.,  2001, ApJ, 552, 738

\bibitem[\protect\astroncite{{Coburn}, {Rothschild} \&
  {Heindl}}{2000}]{cob00iauc}
{Coburn}, W., {Rothschild}, R., \& {Heindl}, W.~A.,  2000, IAU~Circular, 7487,
  1

\bibitem[\protect\astroncite{{Cook} \& {Page}}{1987}]{cook87}
{Cook}, M.~C., \& {Page}, C.~G.,  1987, MNRAS, 225, 381

\bibitem[\protect\astroncite{{Corbet}, {Woo} \& {Nagase}}{1993}]{cor93}
{Corbet}, R. H.~D., {Woo}, J.~W., \& {Nagase}, F.,  1993, A\&A, 276, 52

\bibitem[\protect\astroncite{{Crampton}, {Hutchings} \& {Cowley}}{1978}]{cra78}
{Crampton}, D., {Hutchings}, J.~B., \& {Cowley}, A.~P.,  1978, ApJ, 225, L63

\bibitem[\protect\astroncite{{Cusumano} et~al.}{1998}]{cus98}
{Cusumano}, G., {Di Salvo}, T., {Burderi}, L., {Orlandini}, M., {Piraino}, S.,
  {Robba}, N., \& {Santangelo}, A.,  1998, A\&A, 338, L79

\bibitem[\protect\astroncite{{Dal Fiume} et~al.}{2000}]{dal00cospar}
{Dal Fiume}, D., et~al., 2000,
\newblock in Broad Band X-ray Spectra of Cosmic Sources, ed. K. Makishima, L.
  Piro, T. Takahashi,  COSPAR/Pergamon Press),  399

\bibitem[\protect\astroncite{{Davison}, {Watson} \& {Pye}}{1977}]{dav77}
{Davison}, P. J.~N., {Watson}, M.~G., \& {Pye}, J.~P.,  1977, MNRAS, 181, 73P

\bibitem[\protect\astroncite{Deeter, Boynton \& Pravdo}{1981}]{dee81}
Deeter, J.~E., Boynton, P.~E., \& Pravdo, S.~H.,  1981, ApJ, 247, 1003

\bibitem[\protect\astroncite{{Delgado-Mart{\'\i{}}} et~al.}{2001}]{del01}
{Delgado-Mart{\'\i{}}}, H., {Levine}, A.~M., {Pfahl}, E., \& {Rappaport},
  S.~A.,  2001, ApJ, 546, 455

\bibitem[\protect\astroncite{{Di Salvo} et~al.}{1998}]{sal98}
{Di Salvo}, T., Burderi, L., Robba, N.~R., \& Guainazzi, M.,  1998, ApJ, 509,
  897

\bibitem[\protect\astroncite{Doxsey et~al.}{1973}]{dox73}
Doxsey, R., Bradt, H.~V., Levine, A., Murthy, G.~T., Rappaport, S., \& Spada,
  G.,  1973, ApJ, 182, L25

\bibitem[\protect\astroncite{{Elsner}, {Ghosh} \& {Lamb}}{1980}]{els80}
{Elsner}, R.~F., {Ghosh}, P., \& {Lamb}, F.~K.,  1980, ApJ, 241, L155

\bibitem[\protect\astroncite{{Forman}, {Tananbaum} \& {Jones}}{1976}]{for76}
{Forman}, W., {Tananbaum}, H., \& {Jones}, C.,  1976, ApJ, 206, L29

\bibitem[\protect\astroncite{{Giacconi} et~al.}{1971}]{gia71}
{Giacconi}, R., {Kellogg}, E., {Gorenstein}, P., {Gursky}, H., \& {Tananbaum},
  H.,  1971, ApJ, 165, L27

\bibitem[\protect\astroncite{{Giacconi} et~al.}{1974}]{gia74}
{Giacconi}, R., {Murray}, S., {Gursky}, H., {Kellogg}, E., {Schreier}, E.,
  {Matilsky}, T., {Koch}, D., \& {Tananbaum}, H.,  1974, ApJS, 27, 37

\bibitem[\protect\astroncite{Gottwald et~al.}{1991}]{got91}
Gottwald, M., Steinle, H., Graser, U., \& Pietsch, W.,  1991, A\&AS, 89, 367

\bibitem[\protect\astroncite{{Grove} et~al.}{1995}]{gro95}
{Grove}, J.~E., et~al., 1995, ApJ, 438, L25

\bibitem[\protect\astroncite{Gruber et~al.}{2001}]{gru01}
Gruber, D.~E., Heindl, W.~A., Rothschild, R.~E., Coburn, W., Staubert, R.,
  Kreykenbohm, I., \& Wilms, J.,  2001, ApJ, 562, 499

\bibitem[\protect\astroncite{Haberl}{1994}]{hab94}
Haberl, F.,  1994, A\&A, 283, 175

\bibitem[\protect\astroncite{{Haberl} \& {White}}{1990}]{hab90}
{Haberl}, F., \& {White}, N.~E.,  1990, ApJ, 361, 225

\bibitem[\protect\astroncite{{Harding}}{1994}]{har94}
{Harding}, A.~K.,  1994,
\newblock in The Evolution of \xray\ Binaries, ed. S. Holt, C.~S. Day,  (New
  York: American Institute of Physics Press),  429

\bibitem[\protect\astroncite{Harding \& Daugherty}{1991}]{har91}
Harding, A.~K., \& Daugherty, J.~K.,  1991, ApJ, 374, 687

\bibitem[\protect\astroncite{Heindl \& Chakrabarty}{1999}]{biff99}
Heindl, W.~A., \& Chakrabarty, D.,  1999,
\newblock in Proceedings of the Symposium ``Highlights in \xray\ Astronomy in
  honour of Joachim Tr{\"u}mper's 65th birthday'', ed. B. Aschenbach, M.~J.
  Freyberg, Vol. 272,  (Garching: Max-Planck Institute), ~25

\bibitem[\protect\astroncite{{Heindl} \& {Coburn}}{1999}]{hei99iauc}
{Heindl}, W.~A., \& {Coburn}, W.,  1999, IAU~Circular, 7126, 2

\bibitem[\protect\astroncite{Heindl et~al.}{1999a}]{hei99compcrsf}
Heindl, W.~A., et~al., 1999a,
\newblock in The Fifth Compton Symposium, ed. M.~L. McConnell, J.~M. Ryan, Vol.
  366,  (New York: American Institute of Physics Press),  178

\bibitem[\protect\astroncite{Heindl et~al.}{1999b}]{hei99comp0115}
Heindl, W.~A., Coburn, W., Gruber, D.~E., Pelling, M., Rothschild, R.~E.,
  Wilms, J., Pottschmidt, K., \& Staubert, R.,  1999b,
\newblock in The Fifth Compton Symposium, ed. M.~L. McConnell, J.~M. Ryan, Vol.
  366,  (New York: American Institute of Physics Press),  173

\bibitem[\protect\astroncite{{Heindl} et~al.}{1999}]{hei99}
{Heindl}, W.~A., {Coburn}, W., {Gruber}, D.~E., {Pelling}, M.~R., {Rothschild},
  R.~E., {Wilms}, J., {Pottschmidt}, K., \& {Staubert}, R.,  1999, ApJ, 521,
  L49

\bibitem[\protect\astroncite{Heindl et~al.}{2001}]{hei01xtej}
Heindl, W.~A., Coburn, W., Gruber, D.~E., Rothschild, R.~E., Kreykenbohm, I.,
  Wilms, J., \& Staubert, R.,  2001, ApJ, 563, L35

\bibitem[\protect\astroncite{{Hutchings} et~al.}{1979}]{hut79}
{Hutchings}, J.~B., {Cowley}, A.~P., {Crampton}, D., {van Paradus}, J., \&
  {White}, N.~E.,  1979, ApJ, 229, 1079

\bibitem[\protect\astroncite{Illarionov \& Sunyaev}{1975}]{ill75}
Illarionov, A.~F., \& Sunyaev, R.~A.,  1975, A\&A, 39, 185

\bibitem[\protect\astroncite{{Inoue} et~al.}{1984}]{ino84}
{Inoue}, H., {Ogawara}, Y., {Waki}, I., {Ohashi}, T., {Hayakawa}, S.,
  {Kunieda}, H., {Nagase}, F., \& {Tsunemi}, H.,  1984, PASJ, 36, 709

\bibitem[\protect\astroncite{{in't Zand}, {Baykal} \&
  {Strohmayer}}{1998}]{zan98}
{in't Zand}, J. J.~M., {Baykal}, A., \& {Strohmayer}, T.~E.,  1998, ApJ, 496,
  386

\bibitem[\protect\astroncite{{in't Zand}, {Strohmayer} \&
  {Baykal}}{1997}]{zan97}
{in't Zand}, J. J.~M., {Strohmayer}, T.~E., \& {Baykal}, A.,  1997, ApJ, 479,
  L47

\bibitem[\protect\astroncite{Isenberg, Lamb \& Wang}{1998a}]{ise98b}
Isenberg, M., Lamb, D.~Q., \& Wang, J. C.~L.,  1998a, ApJ, 493, 154

\bibitem[\protect\astroncite{Isenberg, Lamb \& Wang}{1998b}]{ise98}
Isenberg, M., Lamb, D.~Q., \& Wang, J. C.~L.,  1998b, ApJ, 505, 688

\bibitem[\protect\astroncite{{Iye}}{1986}]{iye86}
{Iye}, M.,  1986, PASJ, 38, 463

\bibitem[\protect\astroncite{Jahoda}{2000}]{jah00}
Jahoda, K.,  2000,
\newblock in Rossi2000: Astrophysics with the Rossi \xray\ Timing Explorer

\bibitem[\protect\astroncite{Jahoda et~al.}{1996}]{jah96}
Jahoda, K., Swank, J.~H., Giles, A.~B., Stark, M.~J., Strohmayer, T., Zhang,
  W., \& Morgan, E.~H.,  1996, SPIE, 2808, 59

\bibitem[\protect\astroncite{{Kaper}, {Hammerschlag-Hensberge} \&
  {Zuiderwijk}}{1994}]{kap94}
{Kaper}, L., {Hammerschlag-Hensberge}, G., \& {Zuiderwijk}, E.~J.,  1994, A\&A,
  289, 846

\bibitem[\protect\astroncite{{Kelley} et~al.}{1983}]{kel83cen}
{Kelley}, R.~L., {Rappaport}, S., {Clark}, G.~W., \& {Petro}, L.~D.,  1983,
  ApJ, 268, 790

\bibitem[\protect\astroncite{{Kendziorra} et~al.}{1994}]{ken94}
{Kendziorra}, E., et~al., 1994, A\&A, 291, L31

\bibitem[\protect\astroncite{{Kendziorra} et~al.}{1992}]{ken92}
{Kendziorra}, E., et~al., 1992,
\newblock in The Compton Observatory Science Workshop,  (Goddard Space Flight
  Center: NASA),  217,
\newblock see N92-21874 12-90

\bibitem[\protect\astroncite{{Kii} et~al.}{1986}]{kii86}
{Kii}, T., {Hayakawa}, S., {Nagase}, F., {Ikegami}, T., \& {Kawai}, N.,  1986,
  PASJ, 38, 751

\bibitem[\protect\astroncite{Knight}{1982}]{kni82}
Knight, F.~K.,  1982, ApJ, 260, 538

\bibitem[\protect\astroncite{{Koh} et~al.}{1997}]{koh97}
{Koh}, D.~T., et~al., 1997, ApJ, 479, 933

\bibitem[\protect\astroncite{{Kretschmar} et~al.}{1999}]{kre00comp}
{Kretschmar}, P., {Kreykenbohm}, I., {Wilms}, J., {Staubert}, R., {Heindl},
  W.~A., {Gruber}, D.~E., \& {Rothschild}, R.~E.,  1999,
\newblock in The Fifth Compton Symposium, ed. M.~L. McConnell, J.~M. Ryan, Vol.
  366,  (New York: American Institute of Physics Press),  163

\bibitem[\protect\astroncite{{Kretschmar} et~al.}{1997}]{kre97}
{Kretschmar}, P., et~al., 1997, A\&A, 325, 623

\bibitem[\protect\astroncite{Kreykenbohm et~al.}{2002}]{kre01}
Kreykenbohm, I., Coburn, W., Wilms, J., Kretschmar, P., Staubert, R., Heindl,
  W.~A., \& Rothschild, R.~E.,  2002,
\newblock Phase-resolved Spectroscopy of Vela X-1,
\newblock A\&A in press

\bibitem[\protect\astroncite{Kreykenbohm et~al.}{1999}]{kre99}
Kreykenbohm, I., Kretschmar, P., Wilms, J., Staubert, R., Kendziorra, E.,
  Gruber, D., Heindl, W.~A., \& Rothschild, R.,  1999, A\&A, 341, 141

\bibitem[\protect\astroncite{Krzemi\'{n}ski}{1974}]{krz74}
Krzemi\'{n}ski, W.,  1974, ApJ, 192, L135

\bibitem[\protect\astroncite{{La Barbera} et~al.}{2001}]{bar01}
{La Barbera}, A., {Burderi}, L., {Di Salvo}, T., {Iaria}, R., \& {Robba},
  N.~R.,  2001, ApJ, 553, 375

\bibitem[\protect\astroncite{{Lamb}, {Wang} \& {Wasserman}}{1990}]{lam90}
{Lamb}, D.~Q., {Wang}, J. C.~L., \& {Wasserman}, I.~M.,  1990, ApJ, 363, 670

\bibitem[\protect\astroncite{{Lampton}, {Margon} \& {Bowyer}}{1976}]{lam76}
{Lampton}, M., {Margon}, B., \& {Bowyer}, S.,  1976, ApJ, 208, 177

\bibitem[\protect\astroncite{{Leahy} \& {Matsuoka}}{1990}]{lea90}
{Leahy}, D.~A., \& {Matsuoka}, M.,  1990, ApJ, 355, 627

\bibitem[\protect\astroncite{{Leahy} et~al.}{1989}]{lea89b}
{Leahy}, D.~A., {Matsuoka}, M., {Kawai}, N., \& {Makino}, F.,  1989, MNRAS,
  237, 269

\bibitem[\protect\astroncite{{Levine} et~al.}{1988}]{lev88}
{Levine}, A., {Ma}, C.~P., {McClintock}, J., {Rappaport}, S., {van der Klis},
  M., \& {Verbunt}, F.,  1988, ApJ, 327, 732

\bibitem[\protect\astroncite{{Makishima} et~al.}{1984}]{mak84}
{Makishima}, K., {Kawai}, N., {Koyama}, K., {Shibazaki}, N., {Nagase}, F., \&
  {Nakagawa}, M.,  1984, PASJ, 36, 679

\bibitem[\protect\astroncite{Makishima \& Mihara}{1992}]{mak92}
Makishima, K., \& Mihara, T.,  1992,
\newblock in Frontiers of \xray\ Astronomy (Proc. of the 28th Yamada Conf.),
  ed. Y. Tanaka, K. Koyama,  (Tokyo: Uni. Acad. Press), ~23

\bibitem[\protect\astroncite{{Makishima} et~al.}{1999}]{mak99}
{Makishima}, K., {Mihara}, T., {Nagase}, F., \& {Tanaka}, Y.,  1999, ApJ, 525,
  978

\bibitem[\protect\astroncite{{Maloney} \& {Begelman}}{1997}]{mal97}
{Maloney}, P.~R., \& {Begelman}, M.~C.,  1997, ApJ, 491, L43

\bibitem[\protect\astroncite{{Marshall} \& {Ricketts}}{1980}]{mar80}
{Marshall}, N., \& {Ricketts}, M.~J.,  1980, MNRAS, 193, 7P

\bibitem[\protect\astroncite{Mason et~al.}{1976}]{mas76}
Mason, K.~O., White, N.~E., Sanford, P.~W., Hawkins, F.~J., Drake, J.~F., \&
  York, D.~G.,  1976, MNRAS, 176, 193

\bibitem[\protect\astroncite{Mavromatakis}{1993}]{mav93}
Mavromatakis, F.,  1993, A\&A, 276, 353

\bibitem[\protect\astroncite{M\'{e}sz\'{a}ros}{1992}]{meszaros92}
M\'{e}sz\'{a}ros, P.,  1992,
\newblock High-energy radiation from magnetized neutron stars,
\newblock  (Chicago: University of Chicago Press)

\bibitem[\protect\astroncite{{Middleditch} et~al.}{1981}]{mid81}
{Middleditch}, J., {Mason}, K.~O., {Nelson}, J.~E., \& {White}, N.~E.,  1981,
  ApJ, 244, 1001

\bibitem[\protect\astroncite{Mihara}{1995}]{mih95}
Mihara, T.,  1995,
\newblock {\em Ph.D. thesis\/}, University of Tokyo

\bibitem[\protect\astroncite{Mihara et~al.}{1991}]{mih91}
Mihara, T., Ohashi, T., Makishma, K., Nagase, F., Kitamoto, S., \& Koyama, K.,
  1991, PASJ, 43, 501

\bibitem[\protect\astroncite{{Nagase}}{1989}]{nag89}
{Nagase}, F.,  1989, PASJ, 41, 1

\bibitem[\protect\astroncite{{Nagase} et~al.}{1992}]{nag92}
{Nagase}, F., {Corbet}, R. H.~D., {Day}, C. S.~R., {Inoue}, H., {Takeshima},
  T., {Yoshida}, K., \& {Mihara}, T.,  1992, ApJ, 396, 147

\bibitem[\protect\astroncite{{Nagase} et~al.}{1986}]{nag86}
{Nagase}, F., {Hayakawa}, S., {Sato}, N., {Masai}, K., \& {Inoue}, H.,  1986,
  PASJ, 38, 547

\bibitem[\protect\astroncite{{Ohashi} et~al.}{1984}]{oha84}
{Ohashi}, T., et~al., 1984, PASJ, 36, 699

\bibitem[\protect\astroncite{{Orlandini} et~al.}{1998}]{orl98vela}
{Orlandini}, M., et~al., 1998, A\&A, 332, 121

\bibitem[\protect\astroncite{Orlandini et~al.}{1998}]{orl98}
Orlandini, M., et~al., 1998, ApJ, 500, L163

\bibitem[\protect\astroncite{Orlandini \& Fiume}{2001}]{orl01}
Orlandini, M., \& Fiume, D.~D.,  2001, astro-ph/0107531

\bibitem[\protect\astroncite{{Owens}, {Oosterbroek} \& {Parmar}}{1997}]{owe97}
{Owens}, A., {Oosterbroek}, T., \& {Parmar}, A.~N.,  1997, A\&A, 324, L9

\bibitem[\protect\astroncite{{Pelling} et~al.}{1991}]{pel91}
{Pelling}, M.~R., {Rothschild}, R.~E., {MacDonald}, D.~R., {Hertel}, R., \&
  {Nishiie}, E.,  1991, SPIE Proceedings, 1549, 134

\bibitem[\protect\astroncite{{Petterson}}{1975}]{pet75}
{Petterson}, J.~A.,  1975, ApJ, 201, L61

\bibitem[\protect\astroncite{{Pravdo} et~al.}{1995}]{pra95}
{Pravdo}, S.~H., {Day}, C. S.~R., {Angelini}, L., {Harmon}, B.~A., {Yoshida},
  A., \& {Saraswat}, P.,  1995, ApJ, 454, 872

\bibitem[\protect\astroncite{{Pravdo} \& {Ghosh}}{2001}]{pra01}
{Pravdo}, S.~H., \& {Ghosh}, P.,  2001, ApJ, 554, 383

\bibitem[\protect\astroncite{{Pravdo} et~al.}{1979}]{pra79}
{Pravdo}, S.~H., et~al., 1979, ApJ, 231, 912

\bibitem[\protect\astroncite{Pringle}{1996}]{pri96}
Pringle, J.~E.,  1996, MNRAS, 281, 357

\bibitem[\protect\astroncite{{Rickard}}{1974}]{ric74}
{Rickard}, J.~J.,  1974, ApJ, 189, L113

\bibitem[\protect\astroncite{{Roberts} et~al.}{2001}]{rob01astroph}
{Roberts}, M. S.~E., {Michelson}, P.~F., {Leahy}, D.~A., {Hall}, T.~A.,
  {Finley}, J.~P., {Cominsky}, L.~R., \& {Srinivasen}, R.,  2001,
  astro-ph/0103140, ApJ in press

\bibitem[\protect\astroncite{{Rose} et~al.}{1979}]{ros79}
{Rose}, L.~A., {Marshall}, F.~E., {Holt}, S.~S., {Boldt}, E.~A., {Rothschild},
  R.~E., {Serlemitsos}, P.~J., {Pravdo}, S.~H., \& {Kaluzienski}, L.~J.,  1979,
  ApJ, 231, 919

\bibitem[\protect\astroncite{Rothschild et~al.}{1998}]{rot98}
Rothschild, R.~E., et~al., 1998, ApJ, 496, 538

\bibitem[\protect\astroncite{{Rothschild} \& {Soong}}{1987}]{rot87}
{Rothschild}, R.~E., \& {Soong}, Y.,  1987, ApJ, 315, 154

\bibitem[\protect\astroncite{{Rubin} et~al.}{1994}]{rub94}
{Rubin}, B.~C., et~al., 1994,
\newblock in The Evolution of \xray\ Binaries, ed. S. Holt, C.~S. Day,  (New
  York: American Institute of Physics Press),  455

\bibitem[\protect\astroncite{{Sadakane} et~al.}{1985}]{sad85}
{Sadakane}, K., {Hirata}, R., {Jugaku}, J., {Kondo}, Y., {Matsuoka}, M.,
  {Tanaka}, Y., \& {Hammerschlag-Hensberge}, G.,  1985, ApJ, 288, 284

\bibitem[\protect\astroncite{{Santangelo} et~al.}{1998}]{san98}
{Santangelo}, A., {Del Sordo}, S., {Segreto}, A., {Dal Fiume}, D., {Orlandini},
  M., \& {Piraino}, S.,  1998, A\&A, 340, L55

\bibitem[\protect\astroncite{{Santangelo} et~al.}{1999}]{san99}
{Santangelo}, A., et~al., 1999, ApJ, 523, L85

\bibitem[\protect\astroncite{Santangelo et~al.}{2002}]{san01}
Santangelo, A., Segreto, A., Orlandini, M., Parmar, A.~N., Oosterbroek, T., \&
  Campana, S.,  2002, submitted to ApJ Letters

\bibitem[\protect\astroncite{{Sato} et~al.}{1986}]{sat86}
{Sato}, N., {Nagase}, F., {Kawai}, N., {Kelley}, R.~L., {Rappaport}, S., \&
  {White}, N.~E.,  1986, ApJ, 304, 241

\bibitem[\protect\astroncite{Schandl \& Meyer}{1994}]{sch94}
Schandl, S., \& Meyer, F.,  1994, A\&A, 289, 149

\bibitem[\protect\astroncite{Schlegel et~al.}{1993}]{sch93}
Schlegel, E.~M., et~al., 1993, ApJ, 407, 744

\bibitem[\protect\astroncite{{Schwartz} et~al.}{1980}]{sch80}
{Schwartz}, D.~A., {Griffiths}, R.~E., {Bowyer}, S., {Thorstensen}, J.~R., \&
  {Charles}, P.~A.,  1980, AJ, 85, 549

\bibitem[\protect\astroncite{{Scott}, {Leahy} \& {Wilson}}{2000}]{sco00}
{Scott}, D.~M., {Leahy}, D.~A., \& {Wilson}, R.~B.,  2000, ApJ, 539, 392

\bibitem[\protect\astroncite{Shakura et~al.}{1999}]{sha99}
Shakura, N.~I., Prokhorov, M.~E., Postnov, K.~A., \& Ketsaris, N.~A.,  1999,
  A\&A, 348, 917

\bibitem[\protect\astroncite{Shinoda et~al.}{1990}]{shi90}
Shinoda, K., Kii, T., Mitsuda, K., Nagase, F., Tanaka, Y., Makishima, K., \&
  Shibazaki, N.,  1990, PASJ, 42, L27

\bibitem[\protect\astroncite{{Smith} \& {Takeshima}}{1998}]{smi98}
{Smith}, D.~A., \& {Takeshima}, T.,  1998, IAU~Circular, 7014, 1

\bibitem[\protect\astroncite{{Soong} et~al.}{1990}]{soo90a}
{Soong}, Y., {Gruber}, D.~E., {Peterson}, L.~E., \& {Rothschild}, R.~E.,  1990,
  ApJ, 348, 641

\bibitem[\protect\astroncite{Takeshima \& Chakrabarty}{1998}]{tak98iauc}
Takeshima, T., \& Chakrabarty, D.,  1998, IAU~Circular, 7016, 1

\bibitem[\protect\astroncite{{Tamura} et~al.}{1992}]{tam92}
{Tamura}, K., {Tsunemi}, H., {Kitamoto}, S., {Hayashida}, K., \& {Nagase}, F.,
  1992, ApJ, 389, 676

\bibitem[\protect\astroncite{Tanaka}{1986}]{tan86}
Tanaka, Y.,  1986,
\newblock in IAU Colloq. 89: Radiation Hydrodynamics in Stars and Compact
  Objects, ed. D. Mihalas, K.~H. Winkler,  (New York: Springer),  198

\bibitem[\protect\astroncite{{Tananbaum} et~al.}{1972}]{tan72}
{Tananbaum}, H., {Gursky}, H., {Kellogg}, E.~M., {Levinson}, R., {Schreier},
  E., \& {Giacconi}, R.,  1972, ApJ, 174, L143

\bibitem[\protect\astroncite{Telting et~al.}{1998}]{tel98}
Telting, J.~H., Waters, L. B. F.~M., Roche, P., Boogert, A. C.~A., Clark,
  J.~S., de~Martino, D., \& Persi, P.,  1998, MNRAS, 296, 785

\bibitem[\protect\astroncite{Tr{\"u}mper et~al.}{1978}]{tru78}
Tr{\"u}mper, J., Pietsch, W., Reppin, C., Voges, W., Staubert, R., \&
  Kendziorra, E.,  1978, ApJ, 219, L105

\bibitem[\protect\astroncite{{Unger} et~al.}{1998}]{ung98}
{Unger}, S.~J., {Roche}, P., {Negueruela}, I., {Ringwald}, F.~A., {Lloyd}, C.,
  \& {Coe}, M.~J.,  1998, A\&A, 336, 960

\bibitem[\protect\astroncite{{van Kerkwijk} et~al.}{1995}]{ker95}
{van Kerkwijk}, M.~H., {van Paradijs}, J., {Zuiderwijk}, E.~J.,
  {Hammerschlag-Hensberge}, G., {Kaper}, L., \& {Sterken}, C.,  1995, A\&A,
  303, 483

\bibitem[\protect\astroncite{Voges et~al.}{1982}]{vog82}
Voges, W., Pietsch, W., Reppin, C., Tr{\"u}mper, J., Kendziorra, E., \&
  Staubert, R.,  1982, ApJ, 263, 803

\bibitem[\protect\astroncite{{Wang}}{1981}]{wan81}
{Wang}, Y.-M.,  1981, Space Science Reviews, 30, 341

\bibitem[\protect\astroncite{{Wang} \& {Robnik}}{1982}]{wan82}
{Wang}, Y.-M., \& {Robnik}, M.,  1982, A\&A, 107, 222

\bibitem[\protect\astroncite{{Wheaton} et~al.}{1979}]{whe79}
{Wheaton}, W.~A., et~al., 1979, Nature, 282, 240

\bibitem[\protect\astroncite{{White} et~al.}{1976}]{whi76b}
{White}, N.~E., {Mason}, K.~O., {Huckle}, H.~E., {Charles}, P.~A., \&
  {Sanford}, P.~W.,  1976, ApJ, 209, L119

\bibitem[\protect\astroncite{White, Mason \& Sanford}{1977}]{whi77}
White, N.~E., Mason, K.~O., \& Sanford, P.~W.,  1977, Nature, 267, 229

\bibitem[\protect\astroncite{White et~al.}{1976}]{whi76}
White, N.~E., Mason, K.~O., Sanford, P.~W., \& Murdin, P.,  1976, MNRAS, 176,
  201

\bibitem[\protect\astroncite{{White} \& {Swank}}{1984}]{whi84}
{White}, N.~E., \& {Swank}, J.~H.,  1984, ApJ, 287, 856

\bibitem[\protect\astroncite{White, Swank \& Holt}{1983}]{whi83}
White, N.~E., Swank, J.~H., \& Holt, S.~S.,  1983, ApJ, 270, 711

\bibitem[\protect\astroncite{{Wilms} et~al.}{1999}]{wil99}
{Wilms}, J., {Nowak}, M.~A., {Dove}, J.~B., {Fender}, R.~P., \& {di Matteo},
  T.,  1999, ApJ, 522, 460

\bibitem[\protect\astroncite{{Wilson} et~al.}{1998}]{wil98}
{Wilson}, C.~A., {Finger}, M.~H., {Wilson}, R.~B., \& {Scott}, D.~M.,  1998,
  IAU~Circular, 7014, 2

\bibitem[\protect\astroncite{{Wilson}, {Harmon} \& {Finger}}{1999}]{wil99iauc}
{Wilson}, R.~B., {Harmon}, B.~A., \& {Finger}, M.~H.,  1999, IAU~Circular,
  7116, 1

\end{thebibliography}
\end{document}